# First-order transitions in glasses and melts induced by solid superclusters nucleated and melted by homogeneous nucleation instead of surface melting

**Robert F. Tournier**

*Univ. Grenoble Alpes, CNRS, Grenoble INP*, Institut Néel. 38000 Grenoble, France*

**Institute of Engineering Univ. Grenoble Alpes*

## Graphical abstract:

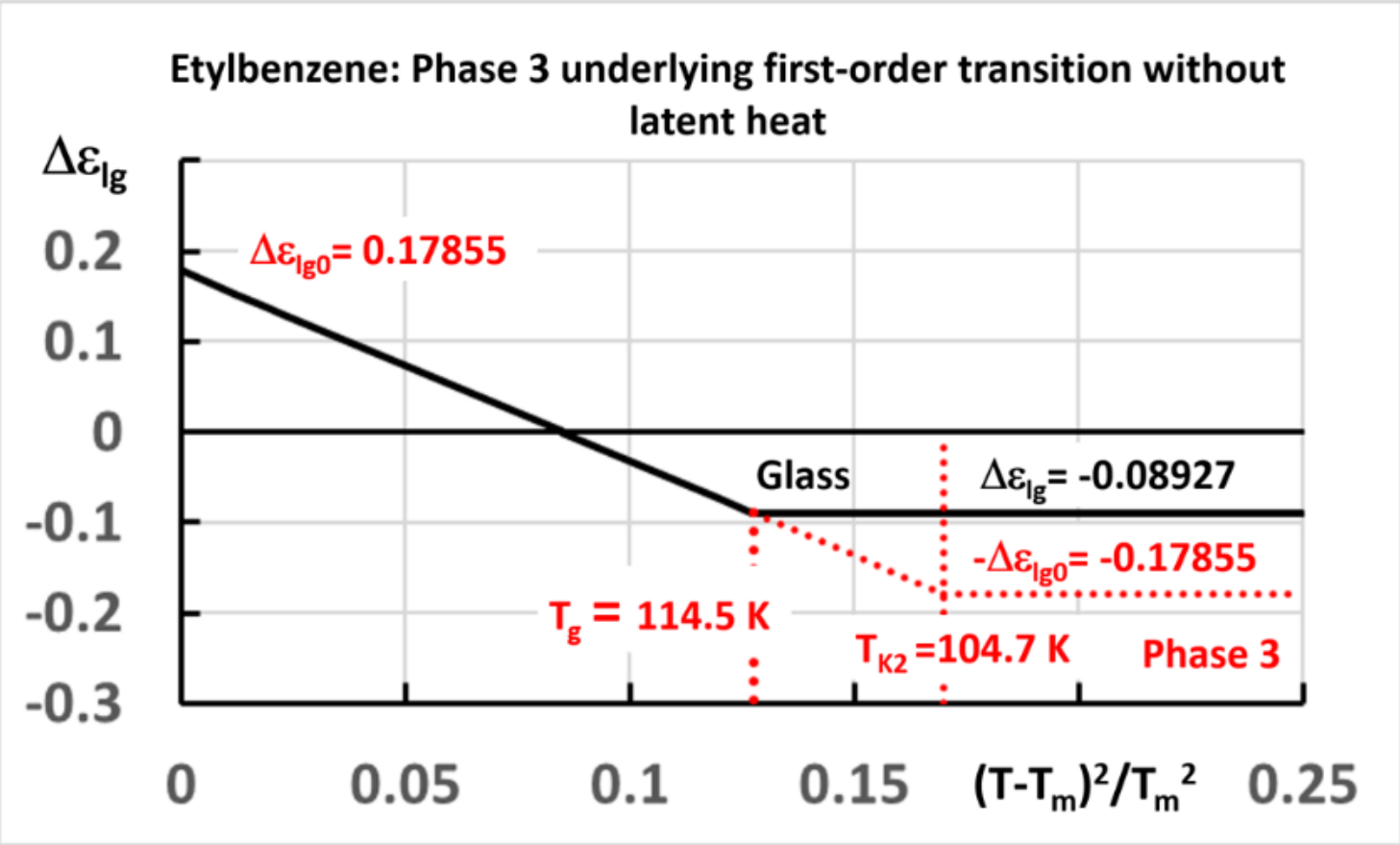

**Ethylbenzene enthalpy $\Delta\varepsilon_{lg}\times\Delta H_m$ below $T_m$ = 178.1 K.**
Undercooled Phase 3 enthalpy coefficient $\Delta\varepsilon_{lg}$ versus $(T-178.1)^2/178.1^2$ in a two-liquid model; $\Delta\varepsilon_{lg}\times\Delta H_m$ being the enthalpy difference between those of Liquid 1 and Liquid 2.
Phase 3 undergoes a first-order transition at $T_{K2}$ = 104.7 K in the absence of glass transition at $T_g$ = 114.5 K and its enthalpy coefficient cannot be lower than (-0.17855).
An underlying first-order transition limits the relaxation enthalpy and fixes the first-order transition temperatures of ultrastable glasses.
Phase 3 undergoes other first-order transitions at higher temperatures associated with glacial, superheated and supercooled phases in various substances.

**Abstract:** Supercooled liquids give rise, by homogeneous nucleation, to solid superclusters acting as building blocks of glass, ultrastable glass, and glacial glass phases before being crystallized. Liquid-to-liquid phase transitions begin to be observed above the melting temperature $T_m$ as well as critical undercooling depending on critical overheating $\Delta T/T_m$. Solid nuclei exist above $T_m$ and melt by homogeneous nucleation of liquid instead of surface melting. The Gibbs free energy change predicted by the classical nucleation equation is completed by an additional enthalpy which stabilize these solid entities during undercooling. A two-liquid model, using this renewed equation, predicts the new homogeneous nucleation temperatures inducing first-order transitions, and the enthalpy and entropy of new liquid and glass phases. These calculations are successfully applied to ethylbenzene, triphenyl phosphite, d-mannitol, n-butanol, $Zr_{41.2}Ti_{13.8}Cu_{12.5}Ni_{10}Be_{22.5}$, $Ti_{34}Zr_{11}Cu_{47}Ni_8$, and $Co_{81.5}B_{18.5}$. A critical supercooling and overheating rate $\Delta T/T_m$ = 0.198 of liquid elements is predicted in agreement with experiments on Sn droplets.

## 1-Introduction

An undercooled liquid develops special clusters, that minimize the energy locally and are incompatible with space filling [1-4]. Such entities are homogeneously formed and act as growth nuclei of crystals when they are crystallized above the glass transition [1,5,6] and below their melting temperature $T_m$. Superclusters containing magic atom numbers are more stable. Their formation temperature out of melt and their radius have been determined for silver [7].

Icosahedral gold nanoclusters do not pre-melt below their bulk melting temperature [8]. The maximum undercooling rate of liquid elements is calculated with an additional enthalpy $\varepsilon_{ls}\times\Delta H_m$ ($\Delta H_m$ being the crystal melting enthalpy) to the Gibbs free energy change predicted by the classical nucleation equation [9] which describes the formation of nuclei under Laplace pressure (called tiny crystals in [10]), and the melting of residual tiny crystals at a second melting temperature above $T_m$ equal to $1.196\times T_m$ [10]. This complementary enthalpy coefficient $\varepsilon_{ls}$ being a linear function of $\theta^2 = (T-T_m)^2/T_m^2$ is at a maximum at $T_m$ and equal to $\varepsilon_{ls0} = 0.217$. This description works, as noted, because the critical size nuclei are melted by liquid homogeneous nucleation instead of surface melting. Other residual superclusters having smaller radii containing magic atom numbers govern the undercooling rate of liquid elements [11]. These last entities melt at much higher temperatures up to $1.3\times T_m$. The value $\varepsilon_{ls0} = 0.217$ of $\varepsilon_{ls}$ at $T_m$ predicts Lindemann's constant of liquid elements equal to 0.103 at $T_m$ [12].

The glass transition at $T_g$ was expected, in the past, to occur at the Vogel-Fulcher-Tammann temperature $T_{VFT}$, which is very close to the Kauzmann temperature $T_{K1}$ of the undercooled liquid phase. The VFT temperature is the temperature at which the viscosity is expected to be infinite. Its value being extrapolated from measurements above $T_g$ is not very precise because $T_g$ is in fact much larger than $T_{K1}$. An underlying first-order transition without latent heat can occur in the supercooled liquid at $T_{K1}$ in the absence of glass transition above $T_{K1}$ and entropy available below $T_{K1}$. An objective of this paper is to determine whether such underlying first-order transition without latent heat occurs at a temperature $T_{K2} > T_{K1}$ and is also hidden by the glass state as inferred by theoretical works [13–16]. In addition, the discovery of ultrastable glass phases obtained by slow vapor deposition at temperatures lower than $T_g$ [17-19] raises the problem of the concomitant existence of a first-order glass-to-glass transition with latent heat at $T_{K2}$ compatible with an undercooled liquid entropy, that is smaller than or equal after transition to $-\Delta S_m$, the crystal entropy.

Two liquid states exist in supercooled water which are separated under pressure and characterized by two values of $T_g$ [20–22]. The glass transition of many melts observed far above $T_{K1}$ also suggests the presence of Liquid 1 and Liquid 2 having different homogeneous nucleation temperatures with two characteristic complementary enthalpy coefficients called $\varepsilon_{ls}$ and $\varepsilon_{gs}$ and two VFT temperatures $T_{0m}$ and $T_{0g}$ ($\varepsilon_{gs0} < \varepsilon_{ls0}$ and $T_{0g} < T_{0m}$) [23,24]. An ordered liquid state viewed as a glass state without enthalpy freezing is expected to occur at the homogeneous nucleation temperature of any fragile liquid. The glass transition at $T_g$ of fragile Liquid 2 is characterized by a phase transition mixing the ordered phases of Liquids 1 and 2 accompanied by an enthalpy change. The two homogeneous nucleation temperatures of strong liquids 1 and 2, being equal, still lead to the mixing of glass phases at $T_g$. The calculated heat capacity jump at $T_g$ and the VFT temperature are in good agreement with the measurements of many samples [24].

The glass transition is now recognized as being a phase transition instead of a liquid freezing. There are other models describing it as a true phase transformation and experimental evidence in favor of this interpretation. The glass transition is seen as a manifestation of critical slowing down near a second-order phase transition with the possible existence of several classes of universality [ (25)]. A model predicting the specific heat jump is based on a percolation-type phase transition with the formation of dynamical fractal structures near the percolation threshold [26–30]. Macroscopic percolating clusters formed at the glass transition have been visualized [31]. High precision measurements of third- and fifth-order non-linear dielectric susceptibilities lead to a fractal dimension $d_F = 3$ for the growing transient domains [32]. An observation of the structural characteristics of medium-range order with neutrons and X-rays leads to $d_F = 2.31$ [33].

The enthalpy difference between Liquid 1 and Liquid 2 depending on $\theta^2$ creates an undercooled phase, that I call Phase 3, which is transformed in glass below $T_g$ and in ordered liquid phase above $T_g$. Phase 3 has two homogeneous nucleation temperatures $T_{n+}$ for melting above $T_m$ and formation below $T_m$ [34]. Its presence has been recognized in supercooled water below $T_m$ and in superheated water under pressure [21,35].

The other objectives of this paper are to extend the application of this renewed equation to glacial phase formation [36-38], glass-to-glass phase transitions such as the expected underlying first-order transitions [14,16], ultra-stable glass formation [18,19], and to the presence of Phase 3 in other melts such as $Ti_{34}Zr_{11}Cu_{47}Ni_8$ [39], $Zr_{41.2}Ti_{13.8}Cu_{12.5}Ni_{10}Be_{22.5}$ (Vit1) **[**40,41**]**, and CoB eutectic alloys [42]. The recent observation of a critical undercooling for crystallization of Sn droplets reveals the critical temperature of solid supercluster formation viewed as growth nuclei when they are crystallized in liquid elements [43]. Superclusters, acting as building blocks in Phase 3, contribute to the ordered liquid formation in glass-forming melts instead of crystallization.

## 2- Predicting glass-to-glass and liquid-to-liquid transitions

### 2.1 Nucleation temperatures of glass Phase 1 and glass Phase 2 in Liquid 1 and Liquid 2

The completed classical nucleation equation for each liquid glass formation [10,11,24].

phase is given by (1):

$$\Delta G = \frac{4\pi R^3}{3}\Delta H_m / V_m \times (\theta - \varepsilon) + 4\pi R^2 (1+\varepsilon)\sigma_1 \qquad (1)$$

where ΔG is the Gibbs free energy change per volume unit, (associated with the formation of a spherical supercluster of radius R), ε is a fraction of the melting enthalpy $\Delta H_m$ per mole (equal to $\varepsilon_{ls}$ for a supercluster in Liquid 1, $\varepsilon_{gs}$ for a supercluster in Liquid 2, and $\Delta\varepsilon_{lg} = (\varepsilon_{ls} - \varepsilon_{gs})$ for a nucleus of Phase 3), $V_m$ is the molar volume, and $\theta = (T-T_m)/T_m$ is the reduced temperature. The melting heat $\Delta H_m$ and the melting temperature $T_m$ are assumed to be the same for all superclusters and not dependent on R, whatever the radius R is. These nuclei are not submitted to surface melting in agreement with earlier findings showing that they are melted by homogeneous nucleation and some of them survive above $T_m$ [10,11]. The critical nuclei give rise to ordered Liquid 1 and ordered Liquid 2 transformed in glass Phase 3 below $\theta_g$, or to various liquid-liquid phase transitions (LLPT), according to the thermal variations of ε. The new surface energy is $(1+\varepsilon) \times \sigma_1$ instead of $\sigma_1$. The classical equation is obtained for $\varepsilon = 0$ [9]. The homogeneous nucleation temperatures $\theta_{n-}$ and $\theta_{n+}$ of these phases are given by (2-3) [24,34]:

$$\theta_{n-} = (\varepsilon-2)/3, \quad (2)$$

$$\theta_{n+} = \varepsilon. \quad (3)$$

The thermally-activated critical barrier is infinite at the homogeneous nucleation temperature obtained for $\theta_{n+} = \varepsilon$ instead of $\theta = 0$ for the classical nucleation equation of crystals [10,24]:

$$\Delta G/kT = 12(1+\varepsilon)^3 Ln(K)/\ 81/\ (\theta-\varepsilon)^2/\ (1+\theta),$$

where $\ln(K) \cong 90$.

A catastrophe of nucleation is predicted at the superheating temperature $\theta_{n+} = \varepsilon$ for crystals protected against surface melting [44]. An ordered liquid phase occurs at $\theta = \theta_{n-}$ in each liquid and disappears by superheating at $\theta = \theta_{n+}$.

The coefficients $\varepsilon_{ls}$ and $\varepsilon_{gs}$ in (4,5) represent values of $\varepsilon(\theta)$, and lead to the nucleus formation having the critical radius in Liquids 1 and 2:

$$\varepsilon_{ls}(\theta) = \varepsilon_{ls0}(1 - \theta^2 \times \theta_{0m}^{-2}), \quad (4)$$

$$\varepsilon_{gs}(\theta) = \varepsilon_{gs0}\left(1 - \theta^2 \times \theta_{0g}^{-2}\right) + \Delta\varepsilon, \quad (5)$$

where $\Delta\varepsilon = |\varepsilon_{ls} - \varepsilon_{gs}|$ is equal to the enthalpy excess coefficient of a quenched liquid before its transition to the glass phase below $T_g$ or the latent heat coefficient $\Delta\varepsilon$ of a first-order transition occurring below $T_g$ as observed for ultrastable glasses or above $T_g$ for glacial phases. The dependence of $\varepsilon$ on $\theta^2$ is deduced from the maximum supercooling rate of 30 pure liquid elements [10]. The enthalpy coefficient $\varepsilon_{ls0} = \varepsilon_{gs0} = 0.217$ of these liquid elements leads to the theoretical value 0.103 of Lindemann's constant [12]. The coefficients $\varepsilon_{ls}$ and $\varepsilon_{gs}$ are equal to zero at the reduced temperatures $\theta_{0m}$ and $\theta_{0g}$ and they correspond to the Vogel-Fulcher-Tammann temperatures named $T_{VFT} = T_{0m}$ of Liquid 1 and $T_{0g}$ of Liquid 2. The viscosity $\eta$ and relaxation time diverge with the decreasing temperature T as shown below:

$$\eta = \eta_0 \exp(B/(T-T_{VFT})$$

The VFT temperatures are the disappearance temperatures of Liquids 1 and 2 enthalpy coefficients $\varepsilon_{ls}$ and $\varepsilon_{gs}$ associated with Liquid 1 and Liquid 2 respectively. There is no more free volume at these temperatures and the viscosity becomes infinite because it is impossible to delocalize any atom [45]. Equations (4-5) are applicable at the homogeneous nucleation temperatures $\theta_{n-}$ given by (2) for glass Phase 1 and glass Phase 2 respectively. The two forms of (6), combining quadratic equations (2) and (5), determine $\theta_{n-}$ for Phase 2:

$$\theta_{n-}^{2}\times\varepsilon_{gs0}\times\theta_{0g}^{-2} + 3\times\theta_{n-} + 2 - \varepsilon_{gs0} - \Delta\varepsilon = 0, \quad (6)$$

$$\varepsilon_{gs0} = (3\theta_{n-} + 2 - \Delta\varepsilon)/(1 - \theta_{n-}^{2}\times\theta_{0g}^{-2}). \quad (6)$$

The solutions for $\theta_{n-}$ are given by (7):

$$\theta_{n-} = (-3 \pm [9 - 4(2 - \varepsilon_{gs0} - \Delta\varepsilon)\varepsilon_{gs0}/\theta_{0g}^{2}]^{1/2})\theta_{0g}^{2}/2\varepsilon_{gs0}, \quad (7)$$

where $\theta_{n-}$ in Liquid 2 for the sign + is viewed as the reduced first-order transition temperature, $\Delta\varepsilon$ is the latent heat coefficient of this first-order transition, and $\theta_{n-}$ for the sign – is the reduced homogeneous nucleation temperature at which the enthalpy excess $\Delta\varepsilon\times\Delta H_m$ starts to be recovered after liquid hyper-quenching at a lower temperature than that of the first-order transition. In the case of a first-order transition due to glacial phase or to ultrastable glass formations, the new glass transition $\theta_g$ occurring at ($\theta_{n-}$) given by (7) depends on the latent heat coefficient $\Delta\varepsilon$.

The two forms of (8) determine the homogeneous nucleation temperature $\theta_{n-}$ for Phase 1, combining (4) at this temperature with (2):

$$\theta_{n-}^{2}\times\varepsilon_{ls0}\times\theta_{0m}^{-2} + 3\times\theta_{n-} + 2 - \varepsilon_{ls0} = 0, \quad (8)$$

$$\varepsilon_{ls0} = (3\theta_{n-} + 2)/(1 - \theta_{n-}^{2}\times\theta_{0m}^{-2}). \quad (8)$$

The reduced homogeneous nucleation temperature $\theta_{n-}$ of Phase 1 in (9) is deduced from (8):

$$\theta_{n-} = (-3 \pm [9 - 4(2 - \varepsilon_{ls0})\varepsilon_{ls0}/\theta_{0m}^{2}]^{1/2})\theta_{0m}^{2}/2\varepsilon_{ls0}, \quad (9)$$

where $\theta_{n-}$ in Eq. (9) is called $\theta_1$ for the sign +. Liquid 1 is ordered by cooling at $\theta_1$ in the no man's land without enthalpy freezing despite a nucleation rate equal to 1. A microscopic structure of elementary superclusters is probably formed below their percolation threshold [46]. The most important glass phase occurs at zero pressure at the homogeneous nucleation temperature $\theta_g = \theta_2$ of Phase 2 in Liquid 2. This transition is accompanied at $\theta_g$ by an enthalpy change from ordered Liquid 1 to glass Phase 3 governed by the difference of enthalpy coefficients $\Delta\varepsilon_{lg} = (\varepsilon_{ls} - \varepsilon_{gs})$ between those of Liquid 1 and Liquid 2. Any strong Liquid 1 has a VFT temperature smaller than or equal to $T_m/3$ while that of any fragile Liquid 1 is higher than $T_m/3$. The two families have different thermodynamic properties below $T_g$. A liquid is fragile when Eq. (6,8) have a double solution [47].

## 2.2 Enthalpy coefficients and specific heat of strong liquids and glasses

The coefficients ($\varepsilon_{gs0}$) in (6) and ($\varepsilon_{ls0}$) in (8), calculated for $\Delta\varepsilon = 0$, are determined from the knowledge of VFT temperatures $\theta_{0g}$, $\theta_{0m}$ and of the reduced glass transition temperature $\theta_g$ respecting $\varepsilon_{ls} = \varepsilon_{gs}$ [24,47]. In the great majority of strong liquids, ($\theta_{0g}$) is equal to -1 because the relaxation time follows an Arrhenius law. The nucleation temperatures $T_1$ and $T_2$ of strong Liquids 1 and 2 are equal to $T_g$.

The specific heat change of strong supercooled liquids is the derivative $d(\varepsilon_{ls}-\varepsilon_{gs})/dT\times\Delta H_m$ given by (10):

$$\Delta C_p(T) = 2\times\theta\times(\varepsilon_{ls0}\times\theta_{0m}^{-2}-\varepsilon_{gs0}\times\theta_{0g}^{-2})\times\Delta H_m/T_m. \quad (10)$$

**2.3 Enthalpy coefficients of fragile liquids and glasses**

The solutions of (6,8) are double for fragile liquids. Values of $\varepsilon_{ls0}$ are given in (11) knowing that $\theta_1 > \theta_g$:

$$\varepsilon_{ls}(\theta = 0) = \varepsilon_{ls0} = 1.5\times\theta_1+2 = a\times\theta_g+2, \quad (11)$$

where a = 1 leads to a well-known specific heat excess ($\Delta C_p (T_g)$) of the supercooled melt at $T_g$ equal to ($1.5\times\Delta H_m/T_m$) [23,24,47,48]. This value a = 1 is deduced from the scaling law followed by the VFT temperature of many polymers [49]. For $a \le 1$, $\Delta C_p (T_g)$ is given in (12):

$$\Delta C_p(T) = 2\times\theta/\theta_g \times\Delta H_m/T_m\times(2.25/a-1.5). \quad (12)$$

The double solution for (8) is obtained when the reduced temperature ($\theta_{0m}$) is given by (13):

$$\theta_{0m}^2 = \frac{8}{9}\varepsilon_{ls0} - \frac{4}{9}\varepsilon_{ls0}^2 . \quad (13)$$

New parameters ($\varepsilon_{gs0}$) and ($\theta_{0g}$) are fixed in Eq. (14,15) and lead to a double solution for (6); ($\varepsilon_{gs0}$) is maximized in (14) and (15) [23,24]:

$$\varepsilon_{gs}(\theta = 0) = \varepsilon_{gs0} = 1.5\times\theta_g+2, \quad (14)$$

$$\theta_{0g}^2 = \frac{8}{9}\varepsilon_{gs0} - \frac{4}{9}\varepsilon_{gs0}^2 . \quad (15)$$

The glass transition reduced temperature occurs at $\theta_g = \theta_2$ and is smaller than that for which $\varepsilon_{ls} = \varepsilon_{gs}$. The reduced temperature $\theta$ where $\Delta\varepsilon_{lg} = 0$ is equal $0.8165\times\theta_g$ for a = 1.

2.4 **Enthalpy coefficient of glass phase and underlying first-order transition without latent heat**

The enthalpy difference coefficient ($\Delta\varepsilon_{lg}$) between Liquid 1 and Liquid 2 in (16) gives rise to the new glass Phase 3 below $\theta_g$ instead of glass Phase 2 and to a new liquid Phase 3 above $\theta_g$ when $\Delta\varepsilon = 0$ [21]:

$$\Delta\varepsilon_{lg}(\theta) = (\varepsilon_{ls} - \varepsilon_{gs}) = \varepsilon_{ls0} - \varepsilon_{gs0} + \Delta\varepsilon - \theta^2\times(\varepsilon_{ls0}/\theta_{0m}{}^2 - \varepsilon_{gs0}/\theta_{0g}{}^2), \qquad (16)$$

where ($\Delta\varepsilon$) is the coefficient of enthalpy excess below $T_g$ being frozen after quenching at the reduced temperature $\theta$ without undergoing a transition to the glass phase and ($\varepsilon_{ls0} - \varepsilon_{gs0} = \Delta\varepsilon_{lg0}$) the melting enthalpy coefficient of Phase 3. ($\Delta\varepsilon$) is equal to $|\Delta\varepsilon_{lg}\,(T)|$ given by (16) below $\theta_g$ without $\Delta\varepsilon$. The temperatures $T_{Br-}$ and $T_{Br+}$ where $\Delta\varepsilon_{lg}(\theta)$ with $\Delta\varepsilon = 0$ is equal to zero are called the branching temperatures of the enthalpy.

It is considered that Phase 3 enthalpy $\Delta\varepsilon_{lg}(\theta)$ in (16) becomes constant for De = 0 below the reduced temperature $\theta_{K2}$ where $\Delta\varepsilon_{lg}(\theta_{K2}) = -\Delta\varepsilon_{lg0} = -(\varepsilon_{ls0} - \varepsilon_{gs0})$. The underlying first-order transition is expected to occur at $\theta_{K2}$. The enthalpy excess De after hyperquenching cannot be larger than the enthalpy coefficient $\Delta\varepsilon_{lg0}$ at $T_m$.

Glass Phase 3, when heated above the glass transition ($\theta_g$), is transformed in liquid Phase 3. This "ordered" liquid can be superheated above $T_g$ and $T_m$ and melted above $T_m$ at the reduced temperature ($\theta_{n+}$) given by (3). Equation (16) is used to calculate $\theta_{n+} = \Delta\varepsilon_{lg}$ in agreement with (3). A new homogeneous nucleation temperature of ordered liquid Phase 3 at a supercooling temperature $\theta_{n+} < 0$ still occurs below $T_m$ for $\theta_{n+} = \Delta\varepsilon_{lg} < 0$. The nucleation of ordered liquid Phase 3 by cooling after overheating and melting has for consequence to replace the nucleation temperature $\theta_1$ of Liquid 1 by $\theta_{n+} = \Delta\varepsilon_{lg} < 0$.

Liquid Phase 3 can be cooled by hyperquenching without glass transition down to the temperature $T_{K2}$ (or $\theta_{K2}$) where $\Delta\varepsilon_{lg}$ attains its minimum value ($-\Delta\varepsilon_{lg0} = -(\varepsilon_{ls0} - \varepsilon_{gs0})$) defined by its enthalpy coefficient $\Delta\varepsilon_{lg0}$ at $T_m$. This transition is underlying when Phase 3 enters the glass state and defines an upper limit $\Delta\varepsilon_{lg0}\times\Delta H_m$ for the enthalpy frozen after quenching the melt below $T_{K2}$ as expected from theoretical considerations [13–16].

### 2.5 First-order transformation temperature ($T_{sg}$) of hyper-quenched Phase 3 in more-stable glasses

The enthalpy excess $\Delta\varepsilon$ in (16) is obtained after quenching the sample at a reduced temperature $\theta$ without transition at $\theta_g$ and is equal to $|\Delta\varepsilon_{lg}\,(\theta)|$ given by (16) for $\Delta\varepsilon_{lg}$. The initial enthalpy after quenching below $T_g$ is equal to that of Liquid 1. Consequently, the enthalpy coefficient $\Delta\varepsilon_{lg}$ of Phase 3 before transformation is always equal to zero at all quenching temperatures. The thermally-activated critical barrier $\Delta G/kT$ for Phase 3 formation, given below, is infinite for $\Delta\varepsilon_{lg} = 0$ [[24]]:

$$\Delta G/kT = 12(1+\Delta\varepsilon_{lg})^3 \mathrm{Ln}(K)/\ 81/\ (\Delta\varepsilon_{lg})^2/\ (1+\theta).$$

A sharp enthalpy difference leading to ultrastable glass Phase 3 through a first-order transition is expected at each temperature of quenching below $T_g$ for which $\Delta\varepsilon_{lg}$ is equal to zero in (16). This phenomenon leads to the formation of more-stable glasses [17-19,50,51]. The transformation temperature ($T_{sg}$) for a stable glass formation given in (17) is :

$$\theta_{sg} = -\left[\frac{\varepsilon_{ls0}-\varepsilon_{gs0}+\Delta\varepsilon}{\varepsilon_{ls0}\theta_{0m}^{-2}-\varepsilon_{gs0}\theta_{0g}^{-2}}\right]^{\frac{1}{2}}. \quad (17)$$

The temperature $T_{sg}$ depends on the value of $\Delta\varepsilon = |\Delta\varepsilon_{lg}|$ after quenching or is the substrate temperature used for vapor deposition. The latent heat associated with this first-order transformation $\Delta\varepsilon_{lg}(\theta_{sg})\times\Delta H_m$ is recovered at a new $\theta_g$. The maximum of enthalpy difference between Phase 3 and the ultrastable glass phase is obtained at $\theta_{sg} = \theta_{K2}$ and is defined by the melting enthalpy of Phase 3 equal to $\Delta\varepsilon_{ls0}\times\Delta H_m$. The enthalpy recovery of this ultrastable Phase 3 occurs at a temperature $\theta_{n-}$ given by (6) with $\Delta\varepsilon = \Delta\varepsilon_{lg0}$ and is expected to be higher than $T_g$. The glass transition at very low heating rates is known as being time- dependent and higher than $T_g$ in ultrastable glasses [38,50].

**2.6 Determination of the Kauzmann temperature from Phase 3 entropy**

The entropy $\Delta S(T)$ of Phase 3 is calculated from the specific heat $d(\Delta\varepsilon_{lg})/dT\times\Delta H_m$ and is given by (18):

$$\Delta S(T) = -2(\varepsilon_{ls0}/\theta_{0m}^2 - \varepsilon_{gs0}/\theta_{0g}^2)\times\Delta S_m\times(T_m - T)/T_m + 2(\varepsilon_{ls0}/\theta_{0m}^2 - \varepsilon_{gs0}/\theta_{0g}^2)\times\Delta S_m \,\mathrm{Ln}\,(T_m/T). \quad (18)$$

The Phase 3 Kauzmann temperature $T_K$ is the temperature where $\Delta S(T_K) = -\Delta H_m/T_m = -\Delta S_m$ [52].

**3- Underlying first-order transition below $T_g$ at zero pressure**

**3.1 Ethylbenzene**

The melting temperature $T_m$, the glass transition temperature $T_g$, its reduced value $\theta_g = (T_g - T_m)/T_m$, the specific heat jump $\Delta C_p(T_g)$ of ethylbenzene and the melting enthalpy $\Delta H_m$ are equal to 178.1 K, 114.5 K, -0.3571, 76 $JK^{-1}mol^{-1}$ and 9170 $Jmol^{-1}$ respectively [53]. The enthalpy coefficients $\varepsilon_{ls}$ of Liquid 1, $\varepsilon_{gs}$ of Liquid 2 and $\Delta\varepsilon_{lg}$ of Phase 3 in (19,20,21) and the square of reduced VFT temperatures $\theta_{0g}^2 = 0.34861$ and $\theta_{0m}^2 = 0.26075$ are calculated using (11,13–15) with a = 1 because $\Delta C_p(T_g)$ is equal to $1.5\times\Delta H_m/T_m = 1.5\times\Delta S_m$, $\Delta S_m$ being the melting entropy:

$$\varepsilon_{ls} = 1.6429\times(1-\theta^2/0.26075), \quad (19)$$

$$\varepsilon_{gs} = 1.46435\times(1-\theta^2/0.34861), \quad (20)$$

$$\Delta\varepsilon_{lg} = 0.17855 - \theta^2\times2.1002. \quad (21)$$

At the melting temperature $T_m$, $\Delta\varepsilon_{lg}$ is maximum and equal to $\Delta\varepsilon_{lg0}$ = 0.17855. The coefficient $\Delta\varepsilon_{lg}$ of the supercooled Phase 3 is represented in Figure 1 as a function of the temperature T. $\Delta\varepsilon_{lg}$ is frozen and equal to (-0.08927) below $T_g$ = 114.5 K after slow cooling. The characteristic temperatures are: $T_{K1}$ = 81.7 K calculated with (18); $T_{0m}$ = 87.15 K with (13); $T_{K2}$ = 104.7 K for $\Delta\varepsilon_{lg} = -\Delta\varepsilon_{lg0}$ = -0.17835 in (16); $T_g$ = 114.5 K; $T_{Br-}$ = 126.2 K for $\Delta\varepsilon_{lg}$ = 0 in (16); $T_{n+}$ = 153.5 K for $\Delta\varepsilon_{lg} = \theta_{n+}$ with (16); $T_m$ =178.1 K; $T_{n+}$ = 202.7 K for $\Delta\varepsilon_{lg} =\theta_{n+}$ with (16); $T_{Br+}$ =230 K for $\Delta\varepsilon_{lg}$ =0 with (16). The liquid-liquid transitions predicted at $T_{n+}$ and $T_{Br+}$ have not been observed up to now.

Liquid Phase 3 can be rapidly quenched down to the temperature $T_{K2}$ = 104.66 K without undergoing glass transition. After quenching at $T_{K2}$ and spontaneous transformation in ultrastable glass phase, ($\Delta\varepsilon_{lg}$) initially equal to zero, becomes equal to ($-2\times\Delta\varepsilon_{lg0}$ = -0.3571) on the green line due to a first-order latent heat coefficient $\Delta\varepsilon$ equal to $\Delta\varepsilon_{lg0}$ in (7). The underlying Phase 3 is represented by black dashed lines below $T_g$. Other first-order transitions with latent heat are expected to start from $\Delta\varepsilon_{lg}$ = 0 at any temperature between $T_{K2}$ and $T_g$ because the relaxation time is small and probably of the order of 100 s at any glass transition temperature. The underlying first-order transition without latent heat occurs at $T_{K2}$ with an enthalpy coefficient change of ($-\Delta\varepsilon_{lg0}$). This glass enthalpy coefficient $\Delta\varepsilon_{lg}$ remains constant below $T_{K2}$.

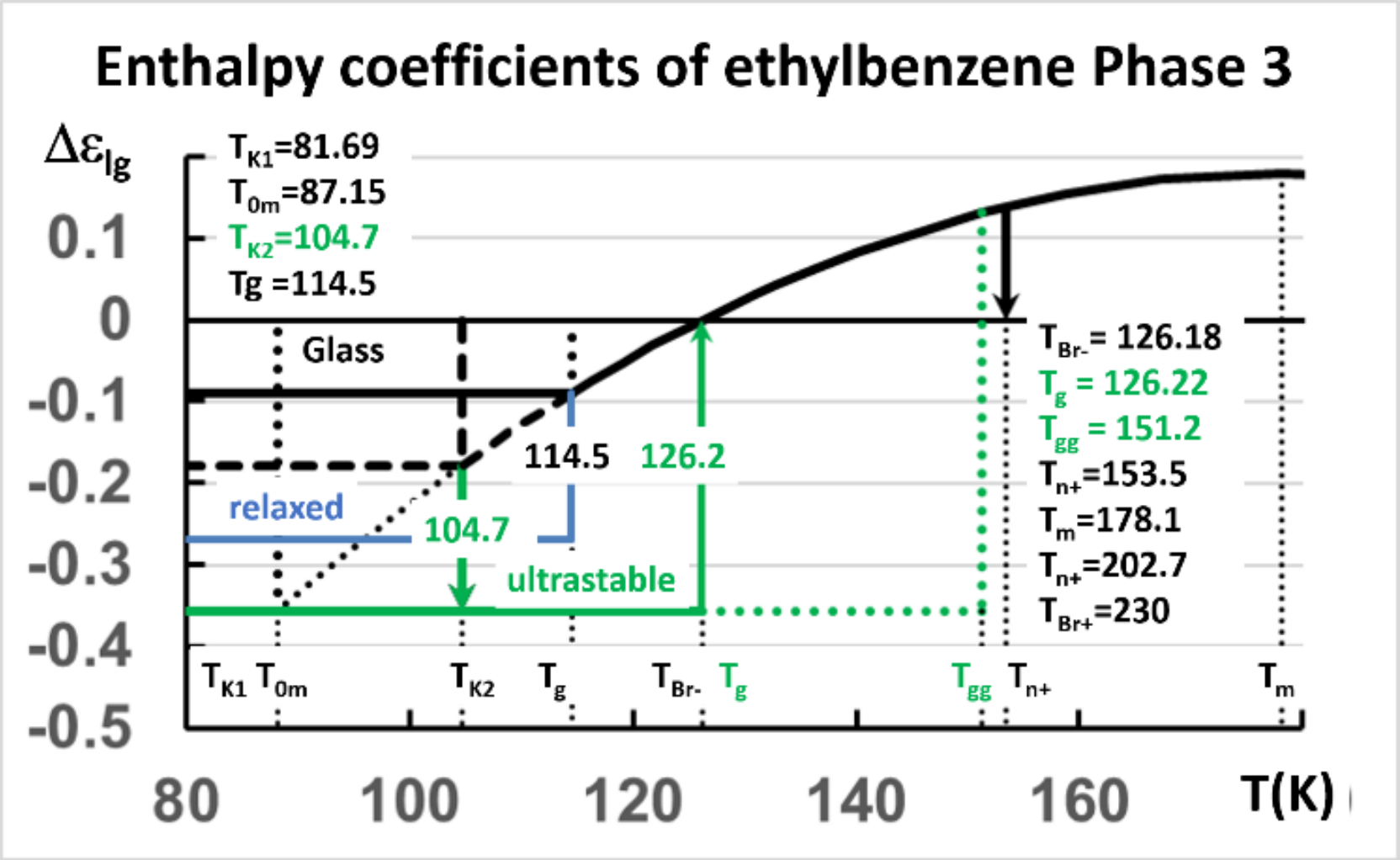

**Figure 1**: Continuous black line: enthalpy coefficient of (–0.08927) for glass phase frozen below $T_g$; black dashed line: enthalpy coefficient $-\Delta\varepsilon_{lg0}$ = -0.17855 of Phase 3 in the absence of glass transition and its underlying first-order transition at $T_{K2}$ without latent heat; blue line: enthalpy coefficient $-1.5\times\Delta\varepsilon_{lg0}$ = -0.2678 of fully-relaxed glass; green line: ultrastable glass phase formation due to the first-order transition at $T_{K2}$ = 104.7 K and its latent heat coefficient $\Delta\varepsilon = \Delta\varepsilon_{lg0}$. Endothermic enthalpy coefficient $-2\times\Delta\varepsilon_{lg0}$ = -0.3571 when recovered at $T_g$ = 126.2 K. Characteristic temperatures of Phase 3: $T_{K1}$ = 81.7 K; $T_{0m}$ = 87.15 K; $T_{K2}$ = 104.7 K; $T_g$ = 114.5 K; $T_{Br-}$ = 126.2 K; $T_{n+}$ = 153.5 K; $T_m$ = 178.1 K; $T_{n+}$ = 202.7 K; $T_{Br+}$ = 230 K.

The glass enthalpy coefficient is composed of a frozen part below $T_g$, obtained by slow cooling, equal to (–0.08927) on the black line instead of ($-\Delta\varepsilon_{lg0}$ = -0.17855) on the dashed black line for supercooled Phase 3 below $T_{K2}$. The irreversible part due the latent heat associated with Phase 3 first-order transition at the reduced temperature θ gives rise to a time-dependent isothermal latent heat recovered at $\theta_g$ which is smaller or equal to ($\Delta\varepsilon_{lg}(\theta)\times\Delta H_m$). This relaxation enthalpy is recovered at $\theta_g$. The enthalpy coefficient of the fully-relaxed glass on the blue line in Figure 1 is ($1.5\times\Delta\varepsilon_{lg0}\times\Delta H_m$ = -0.2678).

The enthalpy variation of ultrastable Phase 3 obtained after quenching or vapor deposition at the first-order transition temperature $T_{K2}$ = 104.7 K is equal to ($2\times\Delta\varepsilon_{lg0}\times\Delta H_m$) on the green line in Figure 1 and recovered at the maximum temperature 126.2 K compatible with the available entropy at this temperature [51]. The glass transition temperature calculated with (7) and Δε = 0.17855 is 151.2 K ($\theta_g$ = -0.15093). The entropy difference of this fragile glass and the liquid calculated with (18), $\varepsilon_{ls}$ and $\varepsilon_{gs}$ with (11-15), and ($\theta_g$ = -0.15093) is equal to the entropy associated with the first-order transition entropy ($0.17855\times\Delta H_m/T_{K2}$) at 126.2 K. The temperature $T_g$ = 126.2 K is the upper limit for the recovery temperature of the enthalpy by isothermal relaxation. A smaller enthalpy variation of ($1.5\times\Delta\varepsilon_{lg}\times\Delta H_m$) is obtained when the enthalpy is recovered at $T_g$ = 114.5 K due to the frozen enthalpy (–0.08927) of this fragile glass [17]. The ultrastable glass Phase 3 induced at $T_{K2}$ has a lower enthalpy than the fully-relaxed glass at the same temperature.

This analysis is confirmed by vapor deposition on substrates cooled at various temperatures [19]. The film volume change is reduced with the increase of the deposition temperature above 104.7 K. The maximum change in Figure 2 is obtained for a deposition temperature T = $T_{K2}$ = 104.7 K as predicted in Figure 1 [51].

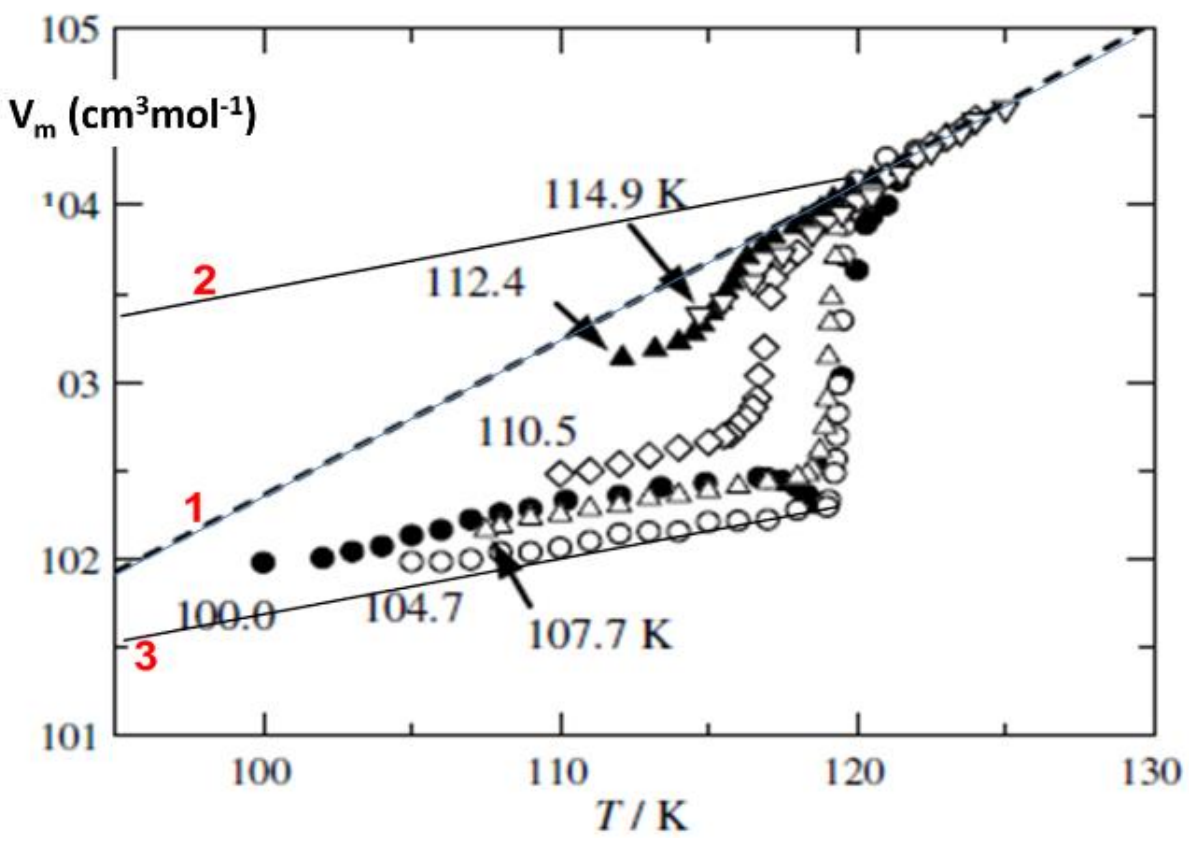

**Figure 2:** Figure already published [19] and completed [51]. Lines 2 and 3 have been added and are parallel to one another. Line 2 represents the molar volume of the liquid below 121 K after

slow cooling. The change in slope between Lines 1 and 2 corresponds to a mean specific heat of 88 J/K/mole. The maximum volume difference is for a deposition temperature of 104.7 K and correspond to a change of enthalpy coefficient of $1.5\times\Delta\varepsilon_{lg0}$.

### 3.2 Enthalpy excess $\Delta\varepsilon_{lg0}\times\Delta H_m$ associated with underlying first-order transitions and first-order transition of ultrastable glass

An underlying first-order transition creates a maximum value $\Delta\varepsilon_{lg0}\times\Delta H_m$ of enthalpy excess which can be frozen by rapid quenching. There are two homogeneous nucleation temperatures defined by (9) in Liquid 2. The highest is $T_{n-} = T_g$; the lowest $T_{n-}$ is the temperature where this enthalpy excess begins to be recovered using a slow heating rate. The enthalpy excesses and the recovery temperatures are examined in 7 glasses. The experimental specific heat jumps at $T_g$ and the recovery temperatures $T_{n-}$ are used to determine $\varepsilon_{ls}$, $\varepsilon_{gs}$, $\Delta\varepsilon_{lg}(T)$, $\Delta C_p(T_g)$, $\Delta\varepsilon_{lg0}\times\Delta H_m$, $\Delta H_m$ and $T_m$ (when they are unknown) in $(CaO)_{55}(SiO_2)_{45}$ [54], $Cu_{46}Zr_{46}Al_8$ [55,56], $Zr_{65}Cu_{27.5}Al_{7.5}$ [57,58], basalt $SiO_2)_{40}(CaO)_{18}(Al_2O_3)_{21}(MgO)_8(FeO)_7(Na_2O)_2(TiO_2)_2(K_2O)_1$ [54], e-glass $(SiO_2)_{55}(CaO)_{17}(Al_2O_3)_{15}(MgO)_5(B_2O_3)_8$ [54]; propylene glycol $C_3H_8O_2$ [59,60], $GeO_2$ [54]. ($T_m$) is known for $(CaO)_{55}(SiO_2)_{45}$, propylene glycol and $GeO_2$. The enthalpy recovery temperature in $GeO_2$ is determined using (17) because the lowest value of $\theta_{n-}$ given by (9) is negative in strong liquids. All the thermodynamic parameters of these glasses are given in Table 1.

**Table 1**: $T_m$ the melting temperature; $T_g$ the glass transition temperature; $\theta_g = (T_g-T_m)/T_m$; $\varepsilon_{gs0}$ ((14)) for fragile liquids and ((6)) for $GeO_2$); $\theta_{0g}^2$ ((15)) for fragile liquids and $\theta_{0g}^2 = 1$ for $GeO_2$; a ((11)); $\varepsilon_{ls0}$ ((11)) for fragile liquids and ((8)) for $GeO_2$); $\theta_{0m}^2$ ((13)) for fragile liquids and $\theta_{0m}^2$ = 0.73381 corresponding to $T_{0m}$ = 200 K for $GeO_2$; $\Delta C_p^{exp}(T_g)$, experimental value; $\Delta C_p^{calc}(T_g)$, calculated value; $H_{exc}^{exp=}$, experimental enthalpy excess; $H_{exc}^{calc}$, calculated enthalpy excess; $\Delta\varepsilon_{lg0} = (\varepsilon_{ls0} - \varepsilon_{gs0})$,; $T_{n-}$ (( 7)); $T_{n-}/T_g$.

| | **$(CaO)_{55}(SiO_2)_{45}$** | **Basalt** | **E-glass** | **$GeO_2$** | **$Cu_{46}Zr_{46}Al_8$** | **$Zr_{65}Cu_{27.5}Al_{7.5}$** | **$C_3H_8O_2$** |
|---|---|---|---|---|---|---|---|
| **$T_m$(K)** | 1790 | 1640 | 1590 | 1388 | 979 | 1110 | 232 |
| **$T_g$(K)** | 1056 | 943 | 962 | 830 | 715 | 660 | 171 |
| **$\theta_g$** | -0.41006 | -0,425 | -0.39497 | -0.40202 | -0.26966 | -0.4054 | -0.2875 |
| **$\varepsilon_{gs0}$** | 1.38492 | 1,3625 | 1.40755 | 0.947 | 1.5955 | 1.39169 | 1.56875 |
| $\theta_{0g}^2$ | 0.3786 | 0.38604 | 0.37062 | 1 | 0.28683 | 0.3769 | 0.30068 |
| **a** | 0.65 | 1 | 1 | | 1 | 1 | 0.875 |
| $\varepsilon_{ls0}$ | 1.67016 | 1.575 | 1.60503 | 1.0182 | 1.7303 | 1.5946 | 1.74844 |
| $\theta_{0m}^2$ | 0.24484 | 0.2975 | 0.28175 | 0.73381 | 0.20738 | 0.28731 | 0.19549 |
| $\Delta C_p{}^{calc}$ $JK^{-1}g^{-1}$ | 0.37 | 0.339 | 0.26 | 0.0408 | 0.167 | 0.219 | 1.02 |
| $\Delta C_p{}^{exp}$ $JK^{-1}g^{-1}$ | 0.37 | 0.34 | 0.26 | 0,045 | 0.16 | 0.19 | 0.986 |
| $H_{exc}{}^{exp}$ $Jg^{-1}$ | 59 | 79 | 54 | 22 | 24 | 32.8 | 25 |
| $H_{exc}{}^{calc}$ $Jg^{-1}$ | 58.9 | 78.8 | 54.3 | 11.37 | 24 | 32.8 | 19.8 |
| $\Delta H_m$ $Jg^{-1}$ | 115 | 371 | 275 | 159.7 | 109.4 | 162 | 110.4 |
| $\Delta\varepsilon_{lg0}$ | 0.16492 | 0.2125 | 0.19748 | 0.0712 | 0.2027 | 0.15145 | 0.1797 |
| $T_{n-}$ | 595 | 541 | 514 | 599 | 521 | 400 | 126.1 |
| $T_{n-}/T_g$ | 0.716 | 0.574 | 0.534 | 0.722 | 0.729 | 0.606 | 0.737 |

The recovery after quenching of the enthalpy excess of propylene glycol is presented in Figure 3 as an example [59]. The recovery starts at $T_{n-}$ = 126 K. The specific heat $\Delta C_p(T_g)$ is calculated with a = 0.875, $T_m$ and $\Delta H_m$ being measured values [60]. The measured enthalpy excess is a little larger than the calculated one. In all other examples, (perhaps except for $GeO_2$), the calculated and measured enthalpy excesses are equal. The enthalpy excess has a maximum value equal to $\Delta\varepsilon_{lg0}\times\Delta H_m$. An underlying first-order transition without latent heat exists and limits the enthalpy excess obtained by quenching.

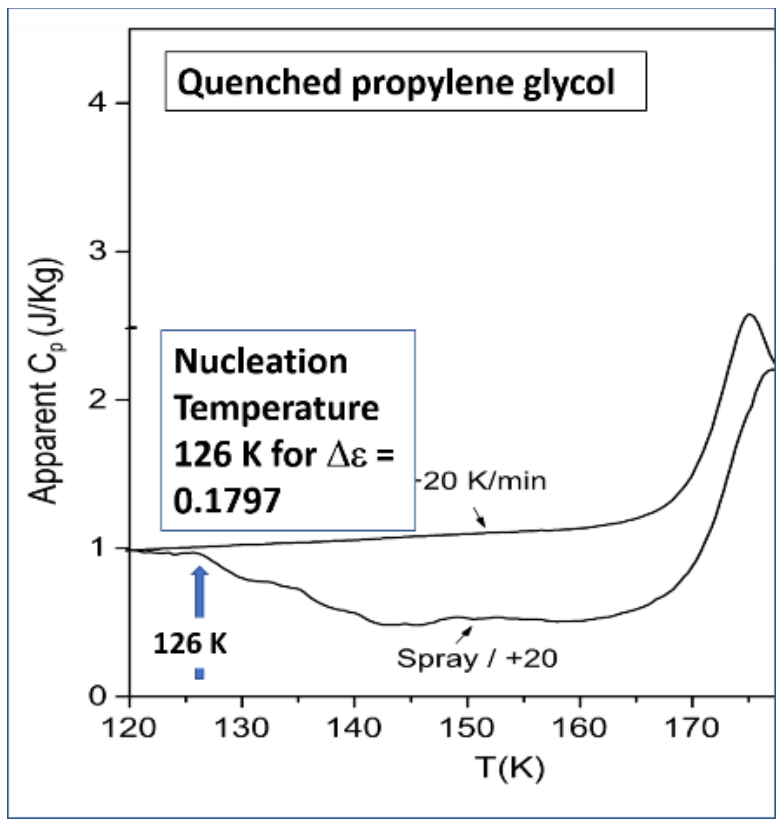

**Figure 3**: The specific heat in $JK^{-1}g^{-1}$ of quenched and slowly-cooled propylene glycol versus T(K) from. The nucleation temperature of 126 K is calculated with (9), $\Delta\varepsilon = 0.1797$, and sign minus. Reproduced from [L.-M. Wang, S. Borick, C.A. Angell. *J. Non-Cryst. Sol.* 353 (2007) 3829-3837] with the permission of Elsevier.

## 4- First-order transitions of glacial phases above $T_g$

### 4.1 Triphenyl phosphite

#### 4.1.1 Specific heat measurements and DSC results

A first-order transition has been discovered in 1995 after an isothermal annealing at 216 K above a glass transition temperature $T_g$ occurring in the range 201.8 to 204 K [36,61–64]. The specific heat jump at $T_g$ is equal to 172.7 $JK^{-1}mol^{-1}$ instead of 124.3 $JK^{-1}mol^{-1} = 1.5\times\Delta H_m/T_m$ for a = 1. It leads to a = 0.8895 in Eq. (12) with $\Delta H_m$ = 25090 $Jmol^{-1}$ and $T_m$ =297 K. The enthalpy differences in Figure 4 between the glass and glacial phases and the glacial and crystalline phases are 7084 and 5744 $Jmol^{-1}$ respectively at 180 K [62]. The glacial phase disappears between 227 and 242 K with a mean $T_{X1} \cong 234.5$ K due to crystallization. These measurements are made using time interval of about 20 min between each specific heat measurement.

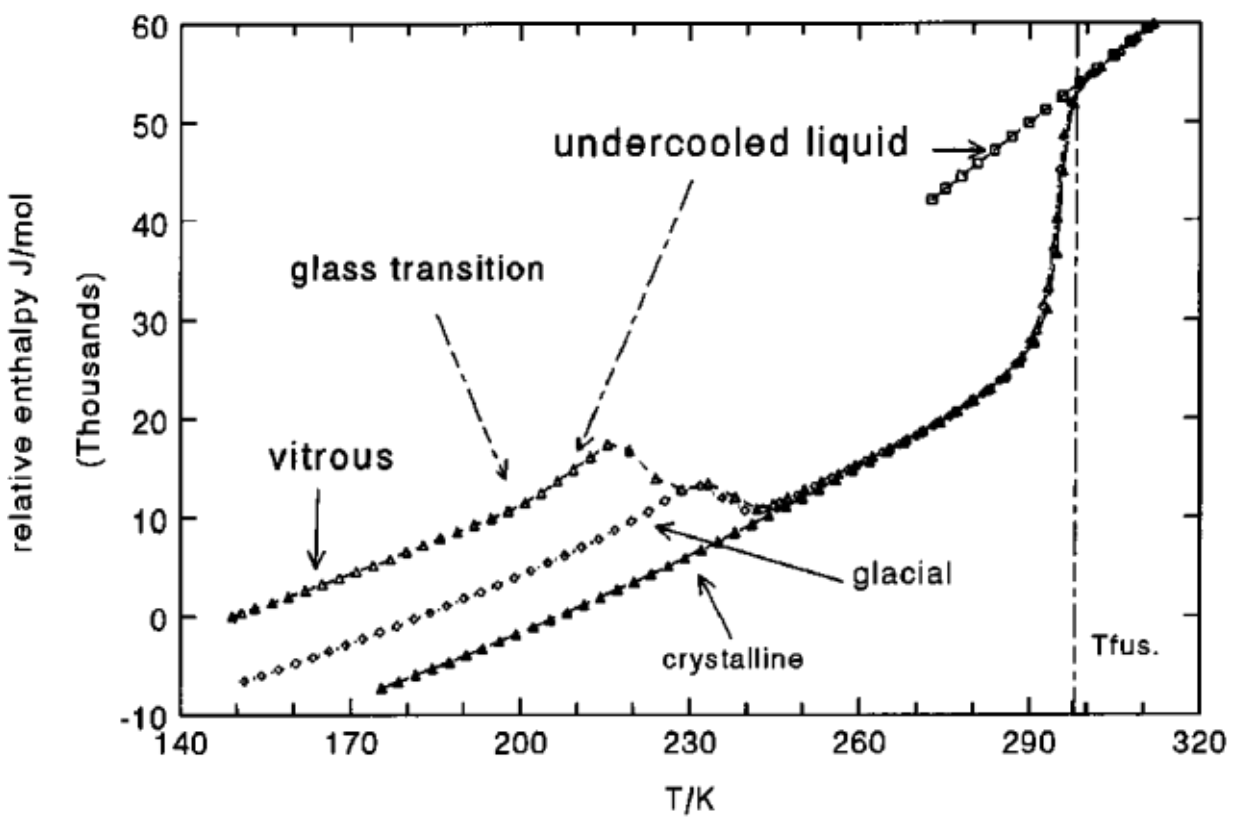

**Figure 4:** Enthalpy values for the different phases. The relative enthalpies have been shifted to overlap at 305 K. Reprinted with permission from K.V. Miltenburg and K. Blok, *J. Phys. Chem.* 100 (1996) 16457-16459. Copyright (1996) Americal Chemical Society

Figure 4 also shows that the glacial phase is vitreous because its enthalpy is frozen up to a temperature of about 227 K where the crystallization begins. Nevertheless, the specific heat increases from 204 K up to 227 K indicating a weak endothermic recovery. The crystallization mainly starts from the glass state. The enthalpy difference between the glass and glacial phases equal to 7084 $Jmol^{-1}$ (22.8 $Jg^{-1}$) results from an isothermal annealing for only 120 min at 217 K and corresponds to an enthalpy coefficient difference equal to 7084/25090 =0.2823.

An extended study of the first-order transition of the glacial phase has recently been published [64]. The heat flow is measured using heating and cooling rates of 1000 $Ks^{-1}$ and reproduced in Figure 5 as a function of temperature before annealing and after an annealing time of 600 min at $T_a$ = 216 K. This work shows the reversibility of the first-order transition induced by isothermal annealing and determines the latent heat of the transition using ultrahigh speed differential scanning calorimetry (DSC). The reverse LLT being hidden behind crystallization is observed in very short times of heating, cooling and reheating cycles as indicated in the inset of Figure 5.

The glass transition temperature $T_g$ increases from 204 to 220 K when varying the heating rate from 5 $Kmin^{-1}$ to 1000 $Ks^{-1}$. The crystallization occurs at $T_{X2}$ for a heating rate of 5 $Kmin^{-1}$. The specific heat at 223 K already shows a bump without applying annealing which corresponds to an endothermic latent heat of about 450 $Jmol^{-1}$ (≅ 0.018 $Jg^{-1}$), which is progressively reduced when the annealing time increases. The melting temperature of the glacial phase is characterized by a peak around $T_{O1}$ = 251 K and a latent heat of about 8690 $Jmol^{-1}$ (28 $Jg^{-1}$) followed by a reverse latent heat of about the same amplitude obtained as indicated in the inset of Figure 5. Crystals are not formed during fast heating after this annealing of 600 min at $T_a$ = 216 K. Higher latent heats are measured when the annealing time and temperature increase, while the reverse latent heat remains equal to 31–33 $Jg^{-1}$, when the first-order latent heat is obtained after an annealing for 600 min from $T_a$ = 218 up to $T_a$ = 227 K (supplementary Figure 2 in [64]). This observation is an evidence of the existence of a critical entropy and enthalpy at $T_{O1}$.

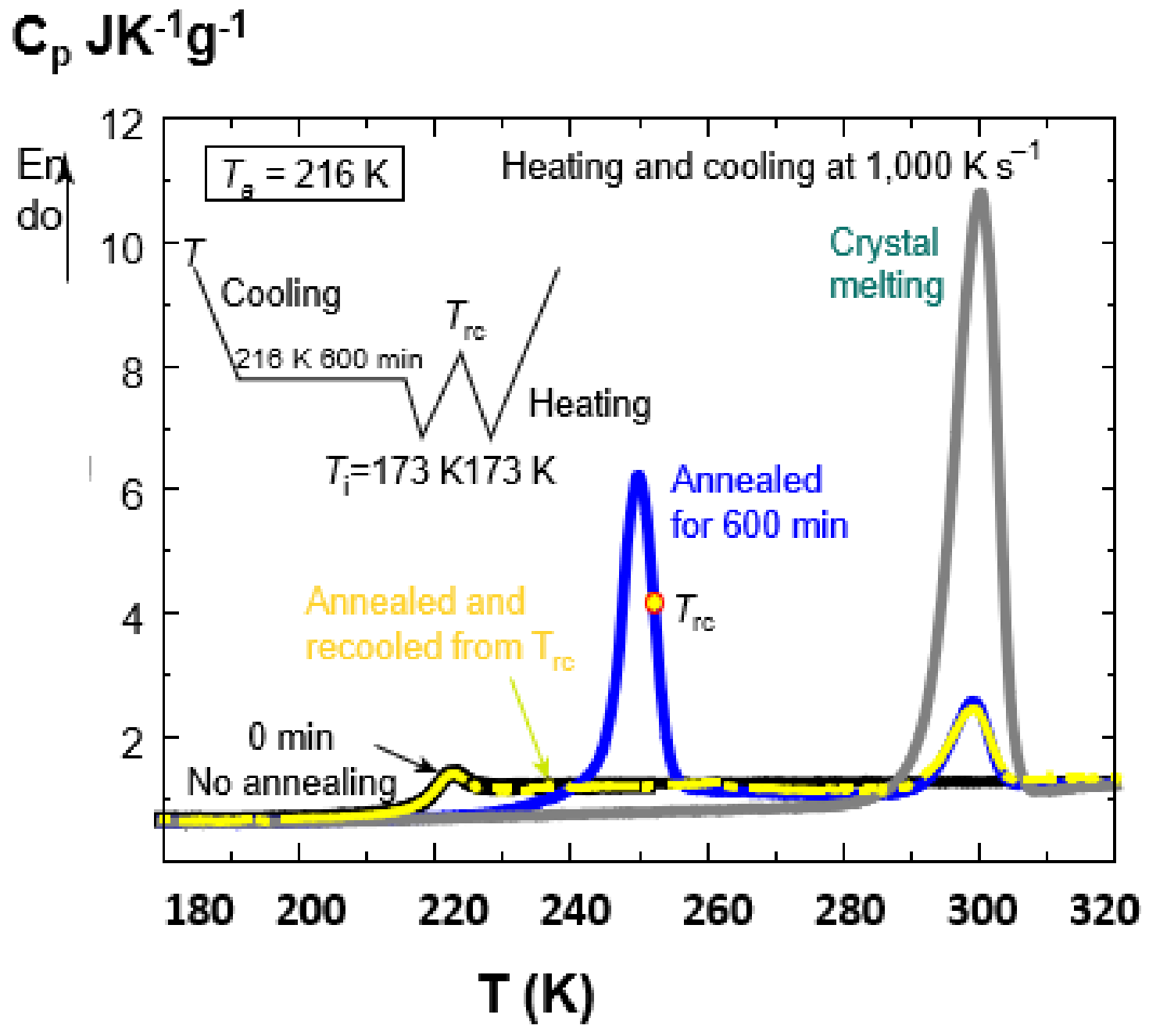

**Figure 5**: Comparison of DSC heat flow curves. The results of flash DSC measurements. The black curve is obtained for a sample without annealing (liquid 1) and the blue curve is for a sample after annealing (the glacial phase, or glass 2). The yellow dashed curve is taken after re-cooled from a point $T_{rc}$ in the endothermic peak. The glass transition signal of liquid 1 is observed in the yellow dashed curve around 220 K, indicating that the glacial phase (glass 2) has already returned to liquid 1 during the endothermic process before reaching $T_{rc}$. The grey curve is for a sample fully crystallized. Reprinted from [M. Kobayashi, & H. Tanaka, *Nature Comm.* 7 (2016) 13438] with the author permission.

### 4.1.2 Model predictions for triphenyl phosphite

The model developed here is used to predict all these phenomena. For that purpose, the enthalpy coefficients of fragile Liquid 1, fragile Liquid 2 and Phase 3 are given in (22-24) using ($\theta_g$ = -0.31313), $T_m$ = 297 K, a = 0.8895 and (11-16):

$$\varepsilon_{ls} = 1.7215\times(1- \theta^2/0.21310), \qquad (22)$$

$$\varepsilon_{gs} = 1.5303\times(1- \theta^2/0.31946), \qquad (23)$$

$$\Delta\varepsilon_{lg} = 0.19117 - \theta^2\times3.2878. \qquad (24)$$

The predicted characteristic temperatures in Figure 6 are in agreement with the experimental observations [62,64] $T_{0m}$ = 159.9 K with (13); $T_{K1}$ = 161.8 K calculated with (18); $T_{K2}$ = 195.7 K with $\Delta\varepsilon_{lg} = -\Delta\varepsilon_{lg0}$ = (-0.19117) in (16); $T_g$ = 204 K; $T_{Br-}$ = 225.4 K for $\Delta\varepsilon_{lg}$ = 0 in (16); the crystallization temperatures $T_{x1}$ = 234.6 K and $T_{x2}$ = 242.5 K determined from entropy considerations; $T_{O1}$ = 253.1 K the calculated first-order transition temperature using entropy considerations with (18); $T_{gg}$ = 280.6 K the virtual glass transition temperature predicted with $\Delta\varepsilon$ = 0.3183 in (7); $T_{n+}$ = 257.5 K and 336.5 K for $\Delta\varepsilon_{lg} = \theta_{n+}$ in (16); $T_m$ = 297 K; $T_{Br+}$ = 368.6 K for $\Delta\varepsilon_{lg}$ =0 in (16). The liquid-to-liquid transitions predicted at $T_{n+}$ and $T_{Br+}$ have not been observed up to now.

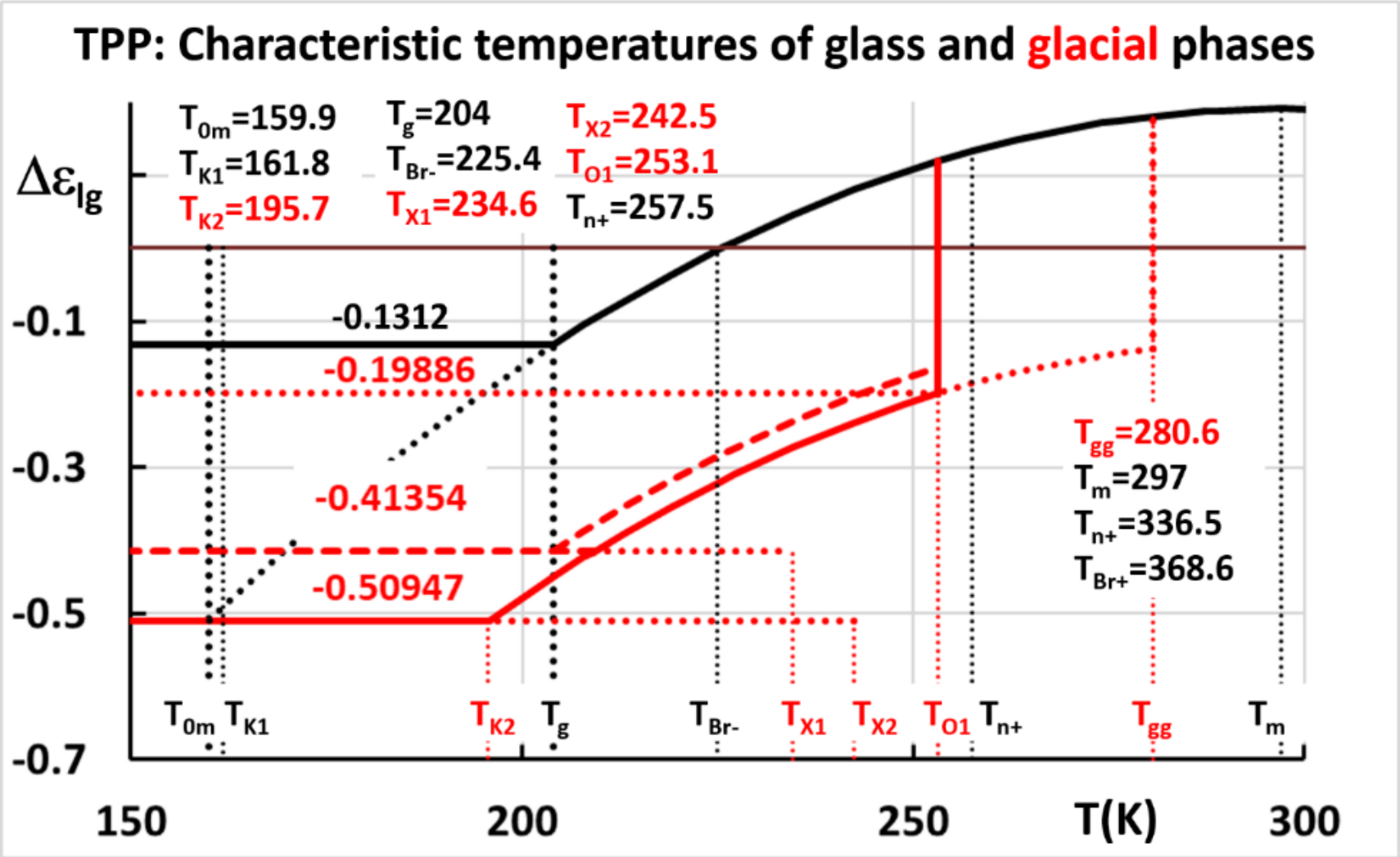

**Figure 6**: Black line: Enthalpy coefficient $\Delta\varepsilon_{lg}$ of Phase 3 with $T_g$ = 204 K. Red lines: enthalpy coefficients of glacial phases equal to $\Delta\varepsilon_{lg}$ of the liquid phase minus $\Delta\varepsilon$ = 0.3183 and $\Delta\varepsilon$ = 0.2823; $T_{K2}$, the underlying first-order transition temperature of Phase 3; $T_{O1}$, the first-order transformation temperatures of glacial phases. Red horizontal lines, the enthalpy coefficients of vitreous glacial phases: (-0.19886) at very high cooling and heating rates; (-0.41354) the enthalpy coefficient of the glacial phase represented in Figure 4 disappearing at $T_{X1}$ (in fact weakly relaxing from 204 to $T_{X1}$ [62]); (-0.50947) the glacial phase enthalpy coefficient at low cooling rates; at $T_{X1}$ and $T_{X2}$, the glass phases crystallize; $T_{gg}$ ((7)), the virtual homogeneous nucleation temperature of the main glacial phase with $\Delta\varepsilon$ = 0.3183 = $(-\Delta\varepsilon_{lg}(T_{0m})$ - $\Delta\varepsilon_{lg0}$ = 0.50947 - 0.19117); at T = $T_{O1}$, the glass and glacial phases return to the liquid state before crystallization.

The enthalpy coefficient of the glacial phase below $T_{K2}$ is chosen as equal to (-0.50947) which is the minimum value of $\Delta\varepsilon_{lg}$ at $T_{0m}$ as shown in Figure 6. This assumption has been already applied to supercooled water to explain the critical enthalpy difference between the two liquid states which are separated under pressure [21]. Values of $\Delta\varepsilon_{lg}$ smaller than (-0.50947) bring the glacial phase closer to the crystal enthalpy. In addition, this glacial phase could be stable. The latent heat of the first-order transition at $T_{O1}$ is equal to (-0.3183) instead of (-0.50947) because it is reduced by the enthalpy change ($-\Delta\varepsilon_{lg0}$ = -0.19117) due to ultrastable glass formation at $T_{K2}$. The latent heat of the first-order transition is predicted to be equal to $0.3183\times\Delta H_m$= 25.74 $Jg^{-1}$ for an experimental value 27.5 $Jg^{-1}$ of the reverse latent heat associated with an annealing for 600 min at 217 K (supplementary Figure 2 in [64]).

The homogeneous nucleation temperatures $T_{n-}$ of superclusters in fragile Liquid 2 are given by (7). There are two nucleation temperatures represented in Figure 7 for a single value of the enthalpy excess $\Delta\varepsilon < 0.19117$; one below $T_g$, another above $T_g$. Values $\Delta\varepsilon < \Delta\varepsilon_{lg0}$ correspond to enthalpy excesses obtained when the cooling rates applied to ribbons are too weak [55]. A coefficient $\Delta\varepsilon$ larger than $\Delta\varepsilon_{lg0}$ cannot lead to a second nucleation temperature $T_{n-}$ below $T_g$ due to the existence of the underlying first-order transition at $T_{K2}$. A coefficient $\Delta\varepsilon = 0.3183$ would have to lead to $T_{n-}$ = $T_{gg}$ = 280.6 K in (7). The entropy of this new fragile liquid with $T_{gg}$ = 280.6 K is calculated using the new coefficients $\varepsilon_{gs0}$ = 1.91697, $\theta_{0g}^2$ = 0.07074, and compared with that of the initial liquid in (18). The transition at $T_{O1}$ = 253.1 K transforms this new glass phase in the initial liquid when the entropy change at $T_{O1}$ becomes equal to $0.3183\times\Delta H_m/T_{O1}$. The enthalpy remains constant below $T_{O1}$ as shown by the systematic reverse enthalpy measurements [64]. ($\Delta\varepsilon_{lg}$) is represented in Figure 6, as being constant and equal to (-0.19886) and also varying, after annealing at $T_a$, instead of being constant below $T_{O1}$.

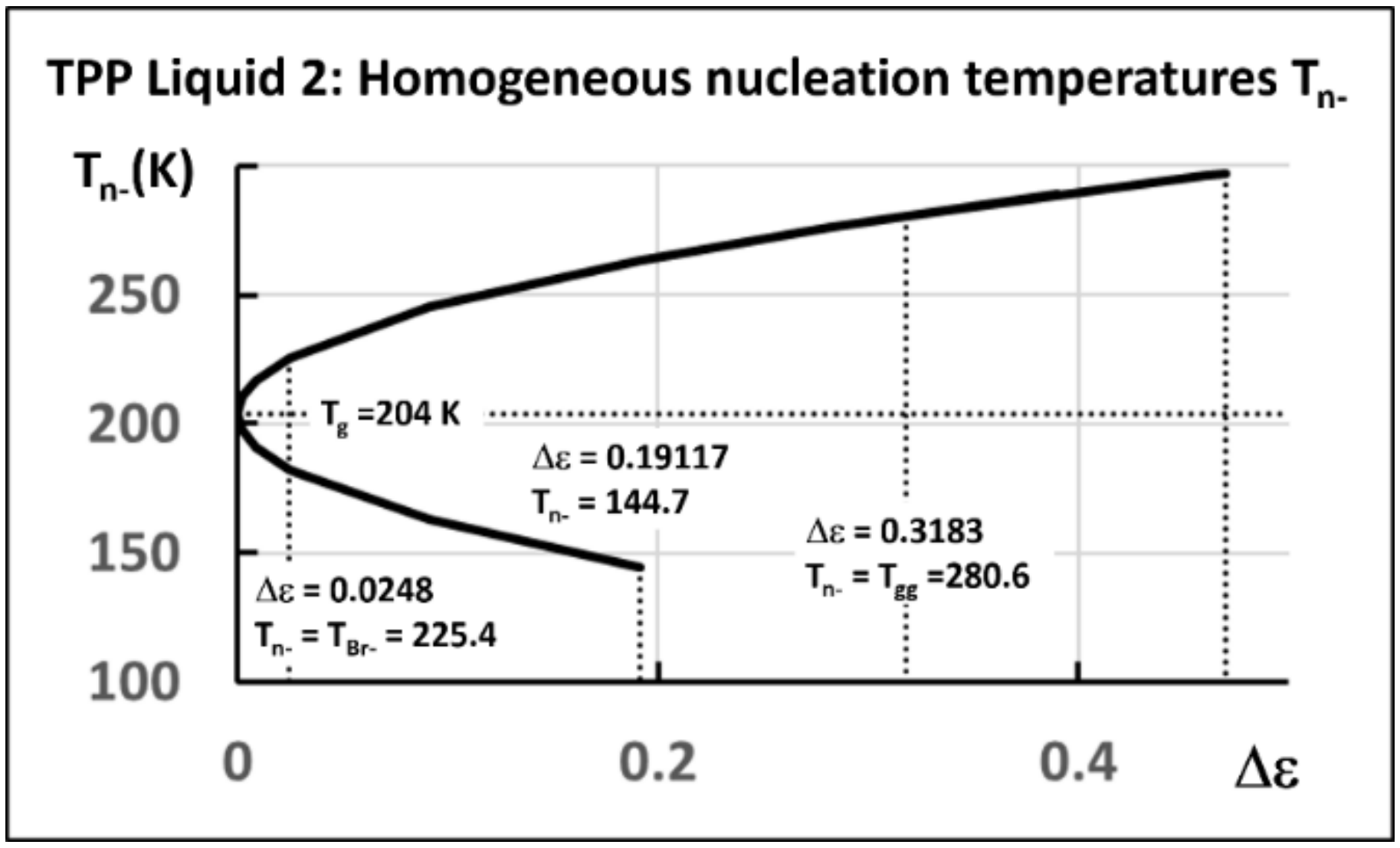

**Figure 7**: Triphenyl phosphite Liquid 2: Homogeneous nucleation temperature $T_{n-}$ with (7) viewed as a glass transition temperature of Liquid 2 versus $\Delta\varepsilon$. For $T_{n-} < T_g$, $(\Delta\varepsilon)$ is an enthalpy excess. For $T > T_g$, $(\Delta\varepsilon)$ is a latent heat coefficient associated with a first-order transition. $(\Delta\varepsilon)$ can also be seen, in the absence of annealing, as a time-dependent endothermic latent heat acting between $T_g = 204$ and $T_{Br-} = 225.4$ K because $\Delta\varepsilon_{lg}$ is still negative in this temperature interval (see Figure 6).

The entropy of glass and glacial phases are calculated using (18) with $\varepsilon_{ls0}$, $\varepsilon_{gs0}$, $\theta_{og}^2$, and $\theta_{om}^2$ given in (22-24), substracting the entropy changes at the first-order transition temperature $T_{O1}$ and using the other calculated characteristic temperatures as represented in Figure 8. The crystallization temperatures $T_{X1}$ and $T_{X2}$ occur when the frozen entropy below $T_{X1}$ and $T_{X2}$ become equal to $\Delta S$ = (-70.22) and (-73.79 $JK^{-1}mol^{-1}$) respectively. These two crystallization temperatures are in perfect agreement with the experimental observations at low heating rates [62,64]. The calculated entropy predicts $T_{X1}$ and $T_{X2}$. The measured latent heat is about 20 % higher than the theoretical one. Consequently, the corresponding frozen entropy would be equal to 81.6 $JK^{-1}mol^{-1}$ instead of 73.79 $JK^{-1}mol^{-1}$. The crystallization would be expected around T = 276 K instead of 242.5 K. The experimental reverse latent heat seems too high to be able to predict $T_{X2}$.

The enthalpy coefficient and entropy of the glacial phase equal to (-0.50947) and (-73.79 $JK^{-1}mol^{-1}$) respectively are obtained by slowly cooling the liquid from the annealing temperature $T_a$ down to $T_{K2}$ = 195.7 K. The liquid phase becomes vitreous with constant enthalpy and entropy up to $T_{X2}$. This description shows that denser glass phases can be obtained using the same process in many other glasses. This analysis is now applied to d-mannitol

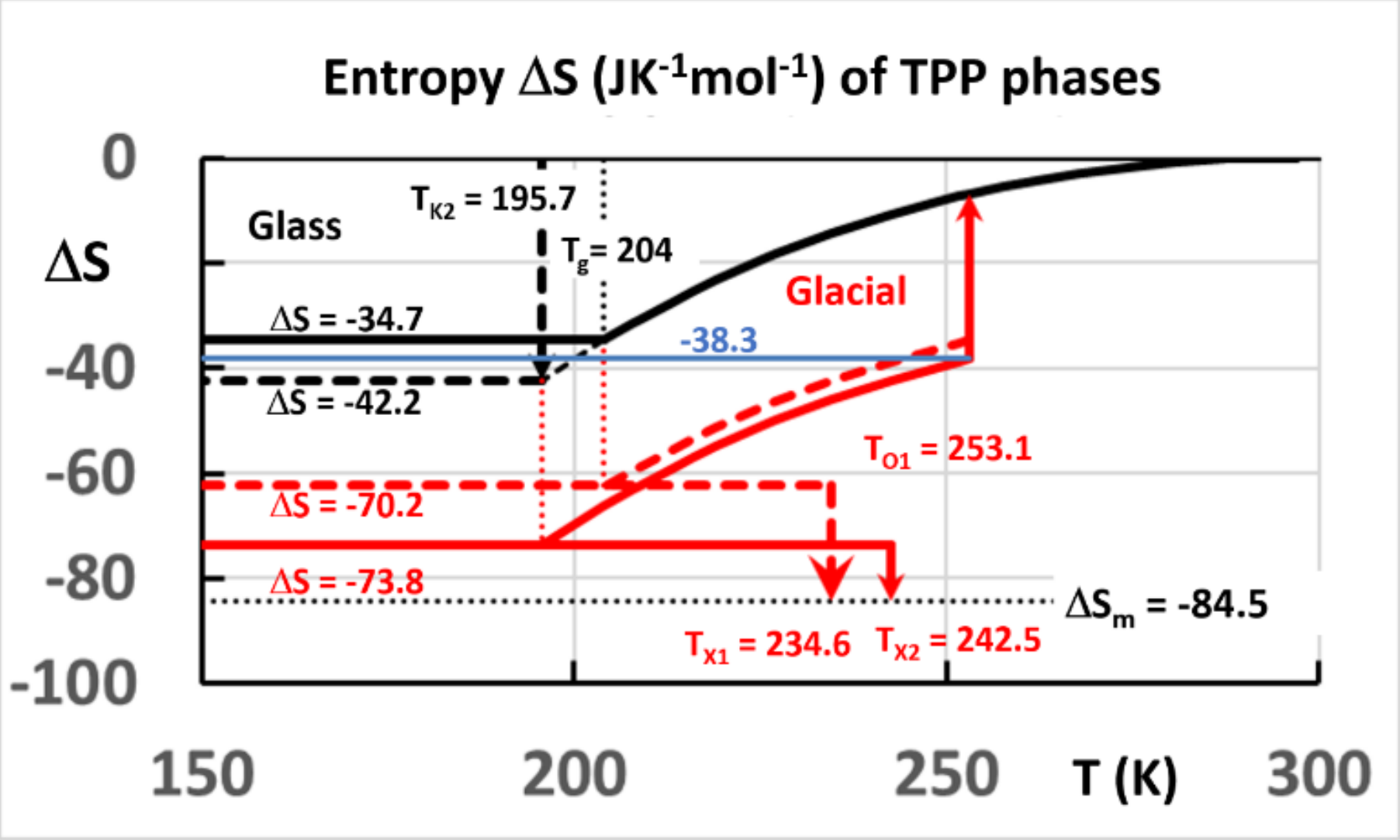

**Figure 8**: Phase 3 entropy in the glass and glacial states. ΔS = (-34.71 $JK^{-1}mol^{-1}$) in the glass state. ΔS = (-38.3 $JK^{-1}mol^{-1}$) in the glacial phase during high rates of cooling and heating corresponding to Δε = 0.3183, ΔS = (-42.24 $JK^{-1}mol^{-1}$) below the underlying first-order transition at $T_{K2}$ =195.7 K. ΔS = (-70.22 $JK^{-1}mol^{-1}$) below $T_{X1}$ = 234.6 K corresponding to the sample annealed for 120 min at $T_a$ = 217 K slowly cooled with Δε = 0.2823 [62] instead of 0.3183. ΔS = (-73.79 $JK^{-1}mol^{-1}$) below $T_{X2}$ =242.5 K. ΔS = (-84.48 $JK^{-1}mol^{-1}$) for the entropy of crystal.

### 4.2 D-mannitol

#### 4.2.1 Specific heat measurements of d-mannitol and DSC results

The glass transition of d-mannitol occurs at $T_g$ = 284 K and the melting temperature at $T_m$ = 439 K. The specific heat jump at $T_g$ is equal to ≈ 1.27 $Jg^{-1}K^{-1}$. The Vogel-Fulcher-Tammann temperature is 222 K [65] and the Kauzmann temperature 229 K. As shown in Figure 9, a spontaneous exothermic heat of 64 $Jg^{-1}$ is produced around 298 K, when heating a sample at 10 $Kmin^{-1}$, initially quenched from the liquid state down to 273 K [38]. This new glacial phase is called Phase X. The crystallization temperature is 331 K and the latent heat of crystallization at this temperature is 107 $Jg^{-1}$. The fusion heat of crystals is equal to 293 $Jg^{-1}$. A fast heating rate of 300 $Ks^{-1}$ reveals an endothermic latent heat of 60 $Jg^{-1}$ at T = 343 K associated with the disappearance of Phase X without crystallization. The glass phase with $T_g$ =284 K could have a higher density at 278 K than the glacial phase at the same temperature. Consequently, these two phases are compared with the HDA and LDA phases of water [66].

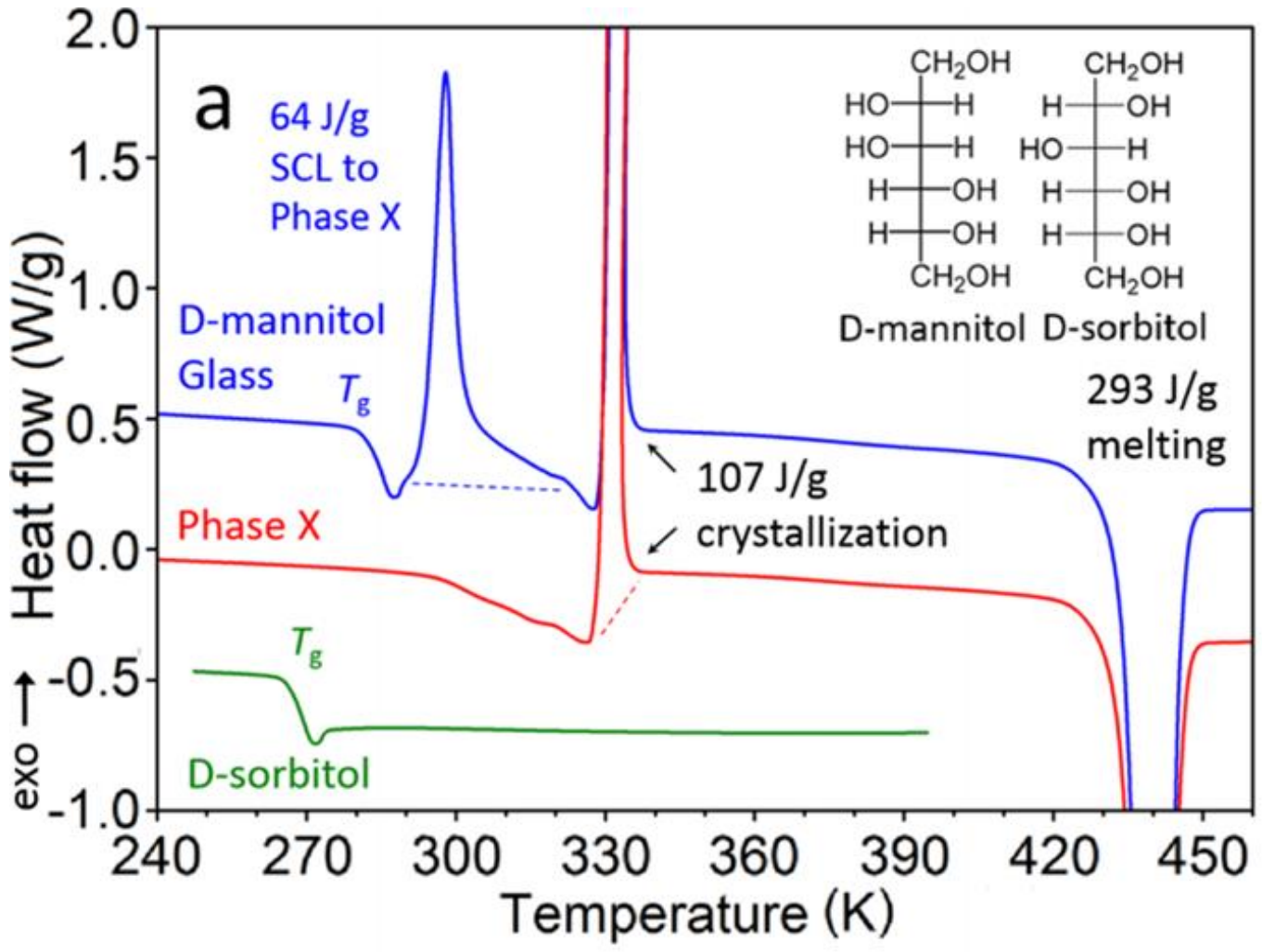

**Figure 9**: DSC traces of the as-prepared glass of d-mannitol, Phase X, and the glass of d-sorbitol. All heating at 10 K/min. Dashed lines indicate baselines for integration. Reproduced from M. Zhu, J-Q Wang, J.H. Perepezko, and L. Yu, *J. Chem. Phys.* 142 (2015) 244504 with the permission of AIP Publishing.

### 4.2.2 Model predictions for d-mannitol

Liquid 1 is fragile because its VFT temperature equal to 222 K is much larger than $T_m/3$ = 146.3 K. Equations (11–15) are applied with $\theta_g$ = (-0.35308) ($T_g$ = 284 K) and a = 0.93 to obtain a specific heat jump of 1.27 $JK^{-1}g^{-1}$ at $T_g$. They lead to ($\varepsilon_{ls}$) of Liquid 1, ($\varepsilon_{gs}$) of Liquid 2, and ($\Delta\varepsilon_{lg}$) of Phase 3 given by (25-27)):

$$\varepsilon_{ls} = 1.67164 \times (1 - \theta^2/0.24396), \quad (25)$$

$$\varepsilon_{gs} = 1.47039 \times \left(1 - \frac{\theta^2}{0.34610}\right), \quad (26)$$

$$\Delta\varepsilon_{lg} = 0.20125 - \theta^2\times 2.6036, \quad (27)$$

where $\Delta\varepsilon_{lg0} = (\varepsilon_{ls0}-\varepsilon_{gs0}) = 0.20125$.

The predicted characteristic temperatures in Figure 10 are in agreement with the experimental observations [38,65,66]: $T_{K1}$ = 220 K calculated with (18); $T_{VFT} = T_{0m}$ = 222.2 K with (13); $T_{K2}$ = 266.4 K with $\Delta\varepsilon_{lg} = \Delta\varepsilon_{lg0}$ = (-0.20115) in (16); $T_g$ = 284 K; $T_a$ = 298.4 K, the spontaneous relaxation temperature of the glacial phase deduced from entropy considerations; $T_{Br-}$ = 317 K for $\Delta\varepsilon_{lg}$ = 0 in (16); the crystallization temperature $T_x$ = 331 K also determined from entropy considerations; $T_{O1}$ = 342.7 K the calculated first-order transition temperature still using entropy considerations with (18); $T_{gg}$ = 382.7 K the virtual glacial transition temperature predicted with $\Delta\varepsilon$ = 0.23872 in (7); $T_{n+}$ = 375 K for $\Delta\varepsilon_{lg} = \theta_{n+}$ in (16); the following temperatures are not indicated in Figure 10: $T_{gg}$ =382.7 K; $T_{n+}$ = 375 and 503 K for $\Delta\varepsilon_{lg} = \theta_{n+}$; $T_m$ = 439 K; $T_{Br+}$ = 561 K for $\Delta\varepsilon_{lg}$ = 0 in (16).

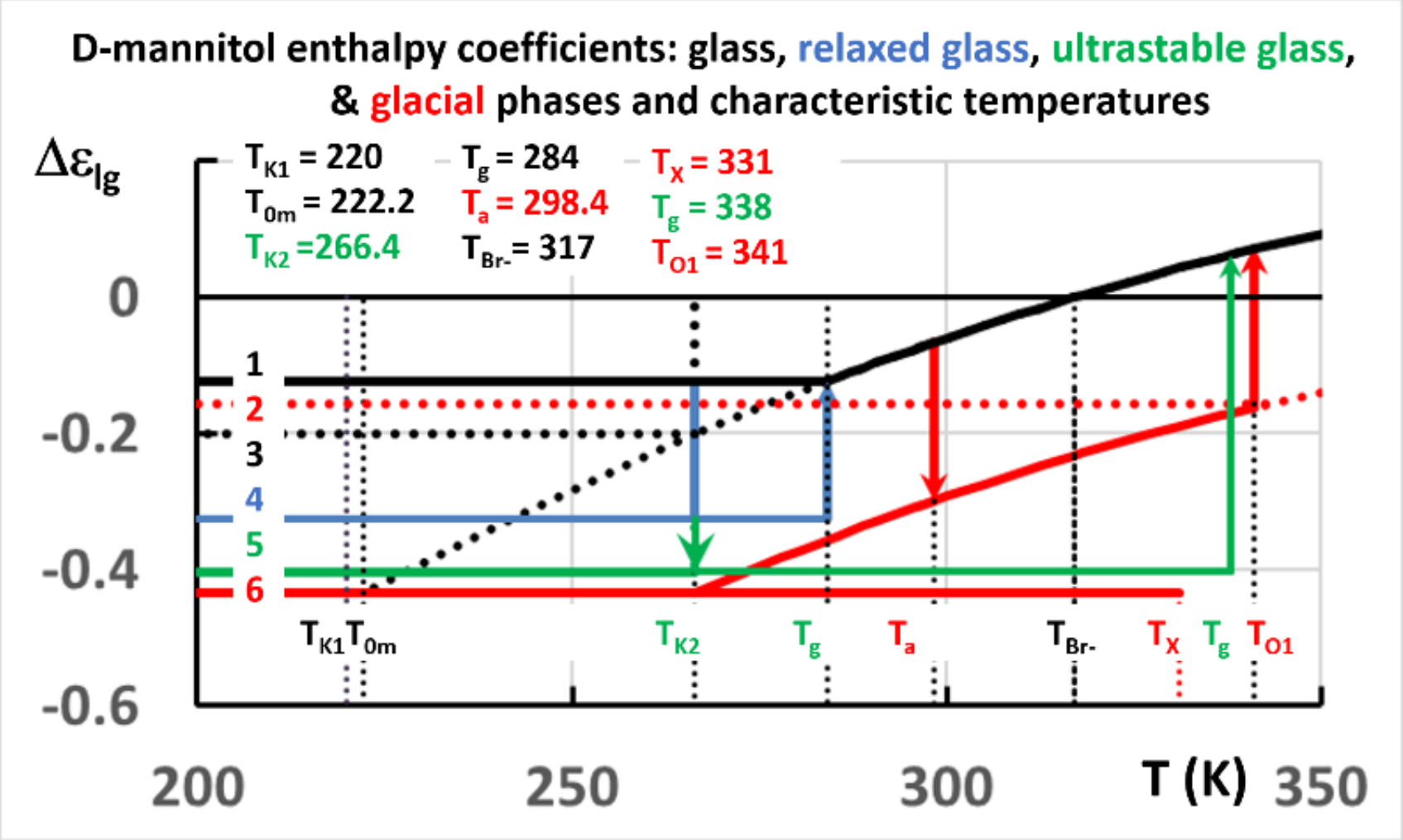

**Figure 10**: D-mannitol: $T_{K1}$ = 220 K, the Kauzmann temperature; $T_{0m}$ = 222.2 K, the VFT temperature; $T_{K2}$ = 266.4 K, the underlying first-order transition temperature; $T_g$ = 284 K, the glass transition temperature; $T_a$ = 298.4 K, the formation temperature of the glacial phase; $T_{Br-}$ = 317 K, the temperature where $\Delta\varepsilon_{lg}$ = 0; $T_X$ = 331 K, the crystallization temperature of the glacial phase; $T_g$ = 338 K, the glass transition temperature of the ultrastable glass phase; $T_{O1}$ = 342.7 K, the first-order transition temperature of the glacial phase. 1- Enthalpy coefficient $\Delta\varepsilon_{lg}$ = (-0.12335) of the glass phase; 2- $\Delta\varepsilon_{lg}$ = (-0.15676) of glacial phase at high cooling and heating rates [38]; 3- $\Delta\varepsilon_{lg}$ = (-0.20126) of undercooled Phase 3 below $T_{K2}$. 4- $\Delta\varepsilon_{lg}$ = (-0.3246) of the fully-relaxed glass phase up to $T_g$ = 284 K. 5- $\Delta\varepsilon_{lg}$ = (-0.40251) of ultrastable Phase 3 formed at $T_{K2}$ = 266.4 K and melted at the upper limit $T_g$ = 338 K. 6- $\Delta\varepsilon_{lg}$ = (-0.43298) of the glacial glass phase during heating at 10 $Kmin^{-1}$ up to the crystallization at $T_X$ = 331 K.

The enthalpy variation of ultrastable Phase 3 obtained after quenching or vapor deposition at the first-order transition temperature $T_{K2}$ = 266.4 K is equal to ($2\times\Delta\varepsilon_{lg0}\times\Delta H_m$) in Figure 10 on the green line 5 and recovered at the maximum temperature 338 K compatible with the available entropy at this temperature. The glass transition temperature calculated with (7) and $\Delta\varepsilon$ = 0.20126 is 379.5 K ($\theta_g$ = -0.13544). The entropy difference of this fragile glass and the liquid calculated with (18), $\varepsilon_{ls}$ and $\varepsilon_{gs}$ with (11-15), and ($\theta_g$ = -0.13144) is equal to the entropy associated with the first-order transition entropy ($0.20126\times\Delta H_m/T_{K2}$) at 338 K. The temperature $T_g$ = 338 K is the upper limit for the recovery temperature of the enthalpy by relaxation. An enthalpy variation of ($1.5\times\Delta\varepsilon_{lg}\times\Delta H_m$) is expected when the enthalpy is recovered at $T_g$ = 284 K due to the frozen enthalpy (–0.12335) of this fragile glass [17]. The ultrastable glass Phase 3 induced at $T_{K2}$ has a lower enthalpy than the fully-relaxed glass at the same temperature represented by the blue line 4.

The homogeneous nucleation temperatures $T_{n-}$ of superclusters in Liquid 2 are given by (7). A coefficient $\Delta\varepsilon$ larger than $\Delta\varepsilon_{lg0}$ cannot lead to a second nucleation temperature $T_{n-}$ below $T_g$ due to the existence of the underlying first-order transition at $T_{K2}$. A coefficient $\Delta\varepsilon = 0.23272$ would have to lead to $T_{n-} = T_{gg} = 382.7$ K in (7) for the glacial phase. The entropy of this new fragile liquid with $T_{gg} = 382.7$ K is calculated using the new coefficients $\varepsilon_{gs0} = 1.80756$, $\theta_{0g}^2 = 0.1546$, and compared with that of the initial liquid in (18). The transition at $T_{O1} = 342.7$ K transforms this new glass phase in the initial liquid when the entropy change at $T_{O1}$ becomes equal to ($0.23872\times\Delta H_m/T_{O1} = 0.2041$ $JK^{-1}$ $g^{-1}$). The enthalpy remains constant below $T_{O1}$ during the heating at 300 $Ks^{-1}$ [38]. ($\Delta\varepsilon_{lg}$) is represented in Figure 10, as being constant and equal to (-0.23872). It also varyies with the heating rate, instead of being the same below $T_{O1}$.

The enthalpy coefficient of the glacial phase (Phase X) below $T_{K2}$ is chosen equal to (-0.43398) which is the minimum value of $\Delta\varepsilon_{lg}$ at $T_{0m}$ as shown in Figure 10. This assumption has already been applied to supercooled water and triphenyl phosphite. The latent heat coefficient of the glacial phase at $T_{O1}$ is added to the first-order latent heat coefficient of ultrastable Phase 3 at $T_{K2}$ equal to (-0.20125). The sum of the two enthalpy coefficients is equal to (-0.43398). That of the first-order transition at $T_{O1}$ is then (-0.23872). The latent heat of the first-order transition is predicted to be equal to $0.23872\times293 = 69.9$ $Jg^{-1}$ in agreement with the experimental values of the formation latent heat 64 $JK^{-1}mol^{-1}$ and the melting latent heat 60 $JK^{-1}mol^{-1}$ [38].

Phase X relaxation at $T_a = 298.4$ K seems to be spontaneous at a heating rate of 10 $Kmin^{-1}$. This phenomenon corresponds to the fact that the liquid entropy up to $T_a$ becomes equal to $\Delta S_m$ when the entropy of the glacial first-order transition at $T_{O1}$ is added to it. $T_a$ is the temperature where the relaxation enthalpy of Phase X becomes available to transform the liquid into a glacial phase.

The entropy of glass and glacial phases are calculated using (18) with $\varepsilon_{ls0}$, $\varepsilon_{gs0}$, $\theta_{og}^2$, and $\theta_{om}^2$ given in (25–27), substracting the entropy changes at the first-order transition temperature $T_{O1}$ for the glacial phase and at $T_{K2}$ for the ultrastable glass phase, using the other calculated characteristic temperatures and represented in Figure 11. The crystallization temperature occurs at $T_X = 331$ K when the frozen entropy below $T_X$ becomes equal to $\Delta S = (-0.5694$ $JK^{-1}g^{-1})$. This crystallization temperature is in perfect agreement with the experimental observation in Figure 10. The calculated entropy is correct because it predicts the experimental value of $T_X$.

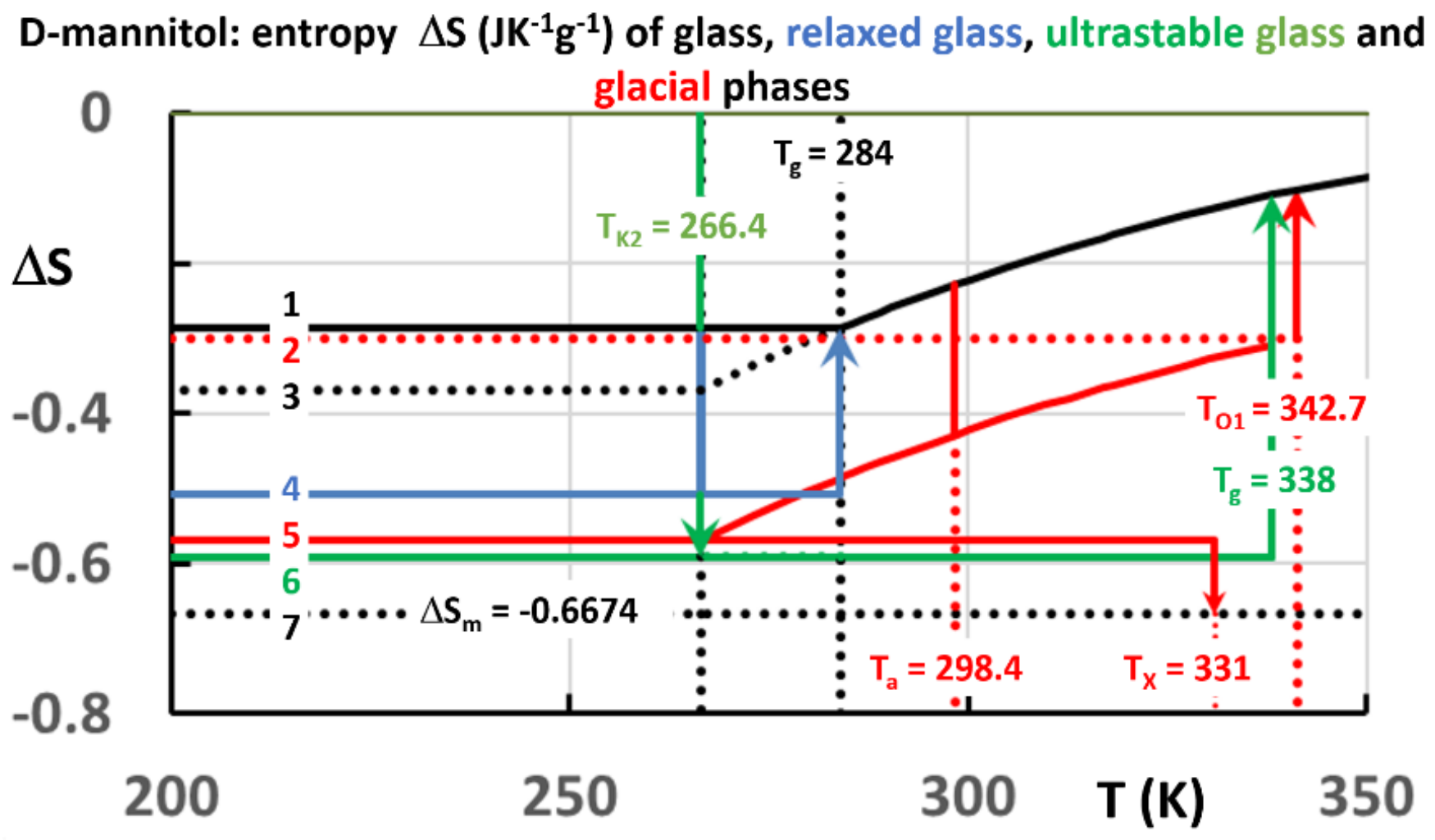

**Figure 11:** D-mannitol: entropy of glass, fully-relaxed glass, ultrastable and glacial phases. 1- Black line: $\Delta S$ = (-0.2865 $JK^{-1}g^{-1}$), glass phase below $T_g$ = 284 K. 2- Red points: $\Delta S$ = (-0.30123 $JK^{-1}g^{-1}$), glacial vitreous phase during heating at 300 $Ks^{-1}$ up to $T_{O1}$ = 341 K. 3- Black points: $\Delta S$ = (-0.36961 $JK^{-1}g^{-1}$), supercooled Phase 3. 4- Blue line: $\Delta S$ = (-0.50784), fully-relaxed glass up to $T_g$ =284 K. 5- Red line: $\Delta S$ = (-0.56938), glacial vitreous phase during slow heating up to the crystallization temperature $T_X$ = 331 K. 6- Green line: $\Delta S$ = (-0.59095 $JK^{-1}g^{-1}$), ultrastable Phase 3 formed at $T_{K2}$ = 266.4 K up to its glass transition at $T_g$ = 338 K. 7- Black points: $\Delta S_m$ = (-0.6674 $JK^{-1}g^{-1}$), crystal.

The entropy (-0.50784 $JK^{-1}g^{-1}$) of fully-relaxed glass phase is recovered at $T_g$ = 284 K as shown in Figure 11 and is composed of the glass entropy (–0.2865 $JK^{-1}g^{-1)}$ and of the relaxation entropy ($-\Delta\varepsilon_{lg0}\times\Delta H_m/T_{K2}$ = -0.22134).

The entropy of ultrastable glass phase formed at $T_{K2}$ = 266.4 K from $\Delta\varepsilon_{lg}$ = 0 is recovered at $T_g$ = 338 K as shown in Figure 11. The first-order transition at $T_{K2}$ is accompanied by an underlying entropy change ($-\Delta\varepsilon_{lg0}$ = -0.36961) and by a first-order entropy change of ($-\Delta\varepsilon_{lg0}\times\Delta H_m/T_{K2}$ = -0.22134) leading to a configurational entropy (-0.59095 $JK^{-1}g^{-1}$).

The enthalpy coefficient and entropy of the glacial phase equal to (-0.43398) and (-0.56939 $JK^{-1}g^{-1}$) respectively are obtained by a slow cooling of the liquid from $T_a$ =298.4 K down to $T_{K2}$ =266.4 K and using a heating of 10 $Ks^{-1}$ up to $T_X$ = 331 K. The liquid phase becomes vitreous with constant enthalpy and entropy up to $T_X$. This description still shows that denser glass phases of bulk samples can be obtained using the same process in many other glasses. This analysis is now applied to n-butanol.

### 4.3 N-Butanol

### 4.3.1 Heat capacity measurements of n-butanol and DSC results

Bolshakov and Dzhonson report in 2005 on the discovery of a new solid phase [67], obtained by isothermal annealing around 140 K, of amorphous n-butanol that melts at 170 K followed by crystallization. The melting temperature of crystals is $T_m$ = 184 K and the melting heat $\Delta H_m$ = 9280 $Jmol^{-1}$, far above the first glass transition temperature $T_g \cong 118$ K [68]. This phenomenon is analogous to the glacial phase formation of triphenyl phosphite [69]. The glass transition at $T \cong 170$ K is associated with a solid amorphous state of n-butanol. Kurita and Tanaka also follow the isothermal transformation of the liquid at a lower temperature 128 K [37]. They observe the formation of many droplets of glacial phase growing in the liquid with time. After 3 hours, the liquid is fully transformed into homogeneous glass having a glass transition $T_g \cong 140$ K as shown in Figure 12. This sample crystallizes at 165 K and melts around 184 K. Hedoux et al. confirm the formation of the glass phase and the progressive appearance of crystallization for annealing times greater than 3 hours at T = 120 K [70]. They evaluate the VFT temperature as being about 45 K, far below $T_m/3$. Consequently, n-butanol is a strong glass. Other calorimetric measurements show that the specific heat jump at $T_g$ is about 48 $JK^{-1}mol^{-1}$, followed by an exothermic peak in the range 125–145 K [71]. The specific heat transition extends from 111 to 118 K. The temperature 116 K at the middle of the transition is chosen to predict the thermodynamic properties with $\Delta H_m$ = 9280 $Jmol^{-1}$.

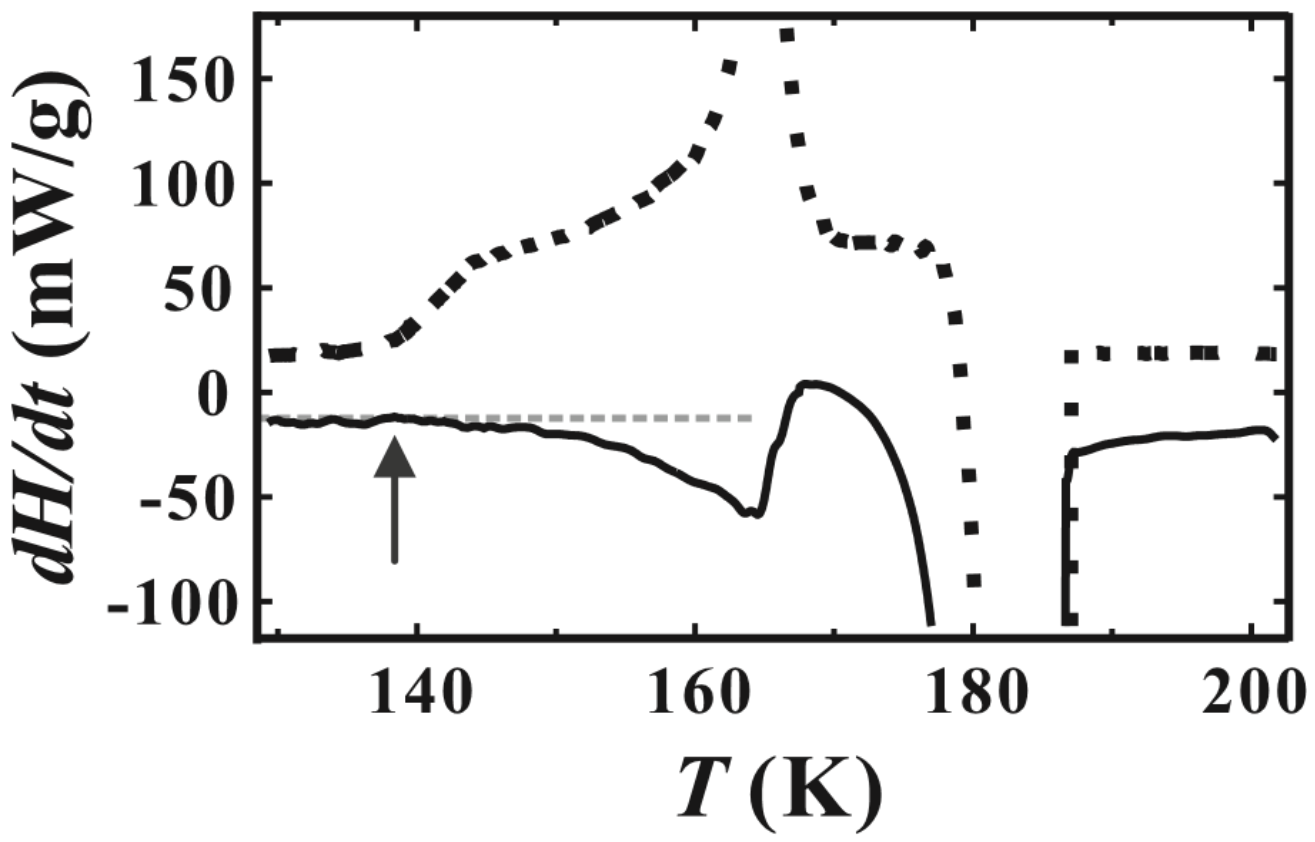

**Figure 12**: Heat flux upon heating across $T_g$ for liquid II of n-butanol. We quenched the sample to 128 K and annealed it for 7 h to completely transform liquid I to liquid II. Then we heated it with AC DSC, where the average heating rate is 3 K $min^{-1}$, the modulation period is 20 s, and the modulation amplitude is 0.16 K. We measured both the reversible part (solid line) and the non-reversible part (dashed line) of the heat flux upon the heating process. The onset of the broad steplike change of the reversible part was observed around 140 K, which is typical of the glass transition. ($T_g^L$) which is the lower edge temperature of the steplike change of the reversible part, was determined to be 140 K. The non-reversible part starts to increase around 140 K. This means

that crystallization starts to occur just after the glass transition. Reprinted from [R. Kurita, and H. Tanaka, *J. Phys.: Condens. Matter*. 17 (2005) L293-L302] with the permission of IOP Publishing.

### 4.3.2 Model predictions for n-butanol

The enthalpy coefficients deduced from (6,8) with $\theta_g$ = (-0.36957) ($T_g$ = 116 K) and $\theta_{0m}^2$ = 0.51546 ($T_{0m}$ = 51.9 K), $\theta_{0g}^2$ = 1 are given in (28–30):

$$\varepsilon_{ls} = 1.2126 \times \left(1 - \frac{\theta^2}{0.51546}\right), \tag{28}$$

$$\varepsilon_{gs} = 1.03229 \times (1 - \theta^2), \tag{29}$$

$$\Delta\varepsilon_{lg} = 0.18030 - \theta^2 \times 1.3202, \tag{30}$$

where $\Delta\varepsilon_{lg0}$ = ($\varepsilon_{ls0}$-$\varepsilon_{gs0}$) = 0.18030. The value $\theta_{0m}^2$ = 0.51546 in (10) leads to $\Delta C_p$ ($T_g$) = 47.9 $JK^{-1}mol^{-1}$ in agreement with specific heat measurements [71] and with the VFT temperature [70].

The predicted characteristic temperatures in Figure 13 are in agreement with the experimental observations [37,68–71]; $T_{K1}$ = 65.5 K calculated with (18); $T_{0m}$ = 51.9 K; $T_{K2}$ = 87.8 K with $\Delta\varepsilon_{lg}$ = $\Delta\varepsilon_{lg0}$ = (-0.18030) in (16); $T_g$ = $T_{Br-}$ = 116 K for $\Delta\varepsilon_{lg}$ = 0 in (16); the crystallization temperature $T_x$ = 165 K determined from entropy consideration; $T_{O1}$ = 140.8 K the calculated first-order transition temperature using (7) with $\Delta\varepsilon$ = 0.31988, $\varepsilon_{gs0}$ =1.03229 and $\theta_{0g}^2$ = 1; $T_{n+}$ = 156.3 for $\Delta\varepsilon_{lg}$ = $\theta_{n+}$ in (16); $T_{O2}$ = 165 K, the first-order transition temperature of the second glacial phase using (7) with $\Delta\varepsilon$ = 0.66911, $\varepsilon_{gs0}$ = 1.03229 and $\theta_{0g}^2$ = 1; $T_m$ = 184 K, the melting temperature. The following temperatures are not indicated in Figure 13: the second $T_{n+}$ = 211.7 K for $\Delta\varepsilon_{lg}$ = $\theta_{n+}$; $T_{Br+}$ = 561 K for $\Delta\varepsilon_{lg}$ = 0 in (16). The liquid-to-liquid transitions predicted at $T_{n+}$ and $T_{Br+}$ have not been observed up to now.

Equation (30) is represented by the black Line 1 as a function of temperature in Figure 13. The enthalpy coefficient of Phase 3 is equal to zero up to $T_g$ = 116 K due the enthalpy freezing below $T_g$ and increases from $T_g$ to $T_m$ where $\Delta\varepsilon_{lg}$ is equal to $\Delta\varepsilon_{lg0}$ = 0.18030. The fully-relaxed glass is represented by the blue line 2. The supercooled Phase 3 without glass transition and underlying first-order transition is represented by a black line of points from 50 K to $T_g$. The underlying first-order transition occurs at $T_{K2}$ = 87.8 K where $\Delta\varepsilon_{lg}$ is equal to (-$\Delta\varepsilon_{lg0}$) and remains constant below $T_{K2}$ along Line 2.

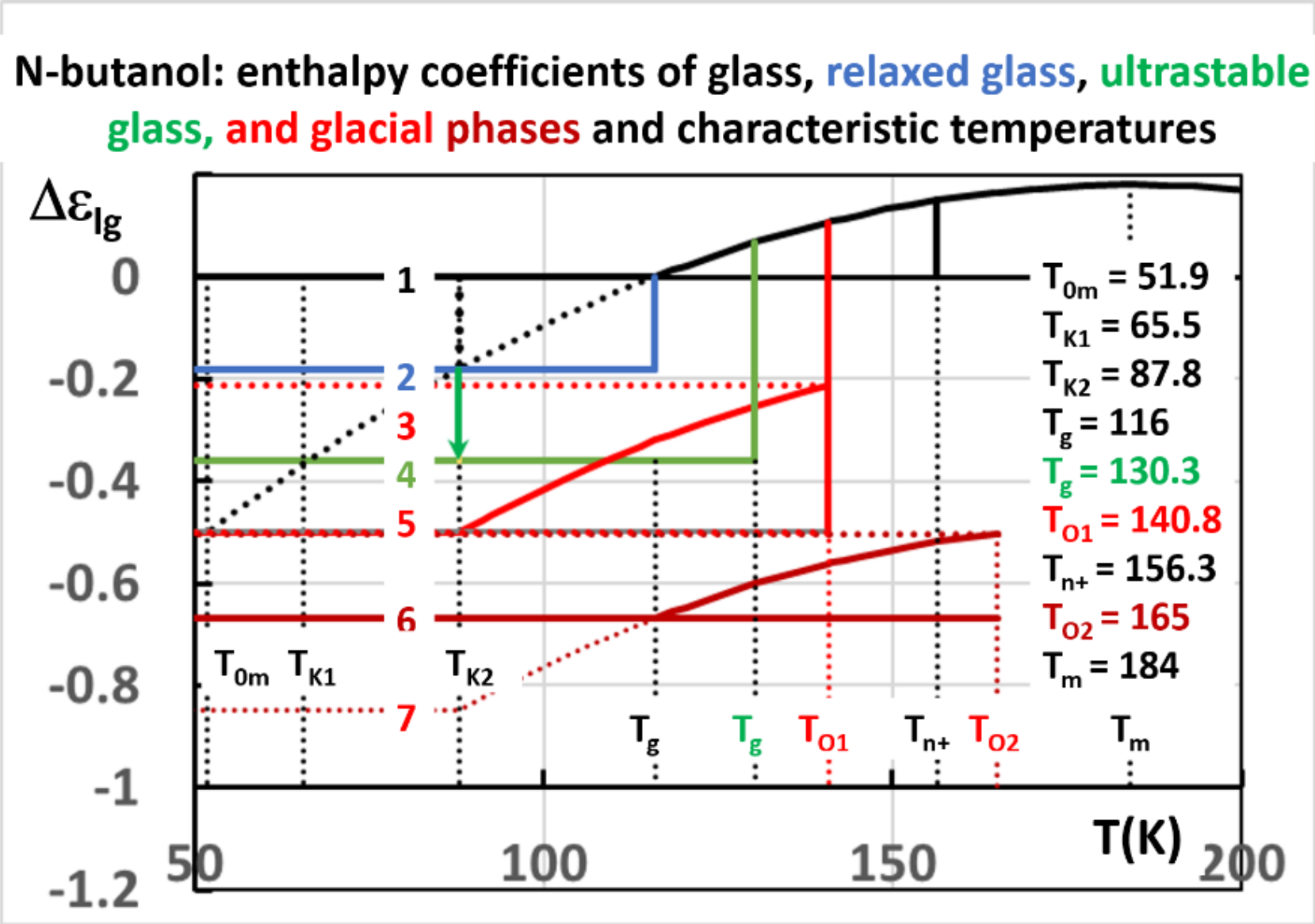

**Figure 13**: N-butanol: enthalpy coefficients $\Delta\varepsilon_{lg}$ of glass, relaxed glass, ultrastable glass phases and two glacial phases versus temperature T(K). 1- $\Delta\varepsilon_{lg}$ = 0, the glass; 2- $\Delta\varepsilon_{lg}$ = (-0.1803), The fully-relaxed glass; 3- $\Delta\varepsilon_{lg}$ = (-0.21244), the first glacial phase at high heating rate and its first-order transition at $T_{O1}$ = 140.8 K; 4- $\Delta\varepsilon_{lg}$ = (-0.3606), the ultrastable glass phase and its glass transition at $T_g$ =130.3 K; 5- $\Delta\varepsilon_{lg}$ = (-0.50018), the first glacial phase at low heating rate, its first-order transition at $T_{O2}$ = 140.8 K and $\Delta\varepsilon_{lg}$ = (-0.50018) at $T_{0m}$ = 51.9 K; 6- $\Delta\varepsilon_{lg}$ = (-0.66911), the second glacial phase from 116 K up to $T_{O2}$ = 165 K also crystallizes at $T_X$ = 165 K; 7- $\Delta\varepsilon_{lg}$ = (-0.84991), the second glacial phase crystallizes below 116 K at low cooling rate (from entropy considerations).

The enthalpy coefficient on the green line 4 in Figure 13 corresponds to the ultrastable glass phase which is formed by the first-order transition of undercooled Phase 3 occurring at $T_{K2}$. The latent heat $\Delta\varepsilon_{lg0}\times\Delta H_m$ is recovered below the upper limit $T_g$ = 130.3 K in the absence of any complementary entropy constraint.

The red Lines 3 and 5 describes the first glacial phase. The enthalpy coefficient of the glacial phase on Line 5 is chosen as equal to the minimum value (-0.50018) of $\Delta\varepsilon_{lg}$ at $T_{0m}$. This assumption has already been applied to supercooled water, triphenyl phosphite, and d-mannitol. The latent heat coefficient of glacial phase at $T_{O1}$ is reduced because there is an underlying first-order transition already at $T_{K2}$ with a coefficient decrease of 0.18030. The sum of the two enthalpy coefficients is assumed to be equal to (-0.50018). That of the first-order transition at $T_{O1}$ is then equal to (-0.31988). The latent heat of the first-order transition is predicted to be equal to (0.31988×$\Delta H_m$ = 2968 $Jmol^{-1}$) with $\Delta H_m$ = 9280 $Jmol^{-1}$. The temperature $T_{O1}$ = 140.8 K calculated using (6) or (7) with $\Delta\varepsilon$ = 0.31988 is in perfect agreement with the observation of this transition in Figure 13. The coefficient $\Delta\varepsilon_{lg}$ is expected to be equal to (-0.21244) along Line 3 up to 140.8 K when the heating rate is high while it is equal to (-0.50018) up to 140.8 K along Line 5 at a heating of 3 $Kmin^{-1}$.

Lines 5, 6 and 7 in Figure 13 characterize the second glacial phase. The glass transition and crystallization occur at $T_{O2}$ = 165 K. This transition is predicted using (7) with Δε = 0.66911 and is observed with a heating of 3 $Kmin^{-1}$. The enthalpy coefficient (-0.66911) is represented on Line 6. The melting of this glass phase occurs along Line 6 at $T_{O2}$ followed by spontaneous crystallization. A much higher heating rate is expected to melt the glass phase at $T_{O2}$ with an enthalpy coefficient equal to (-0.50287) along Line 5 without spontaneous crystallization as observed for triphenyl phosphite [64] and d-mannitol [38]. Line 7 is not attained. The supercooled Phase 3 cannot undergo an underlying first-order transition without crystallization because its frozen entropy would be larger than $\Delta S_m$ = 9280/184 = 50.434 $JK^{-1}$ $mol^{-1}$. This glass phase only exists between $T_g$ and $T_{O2}$.

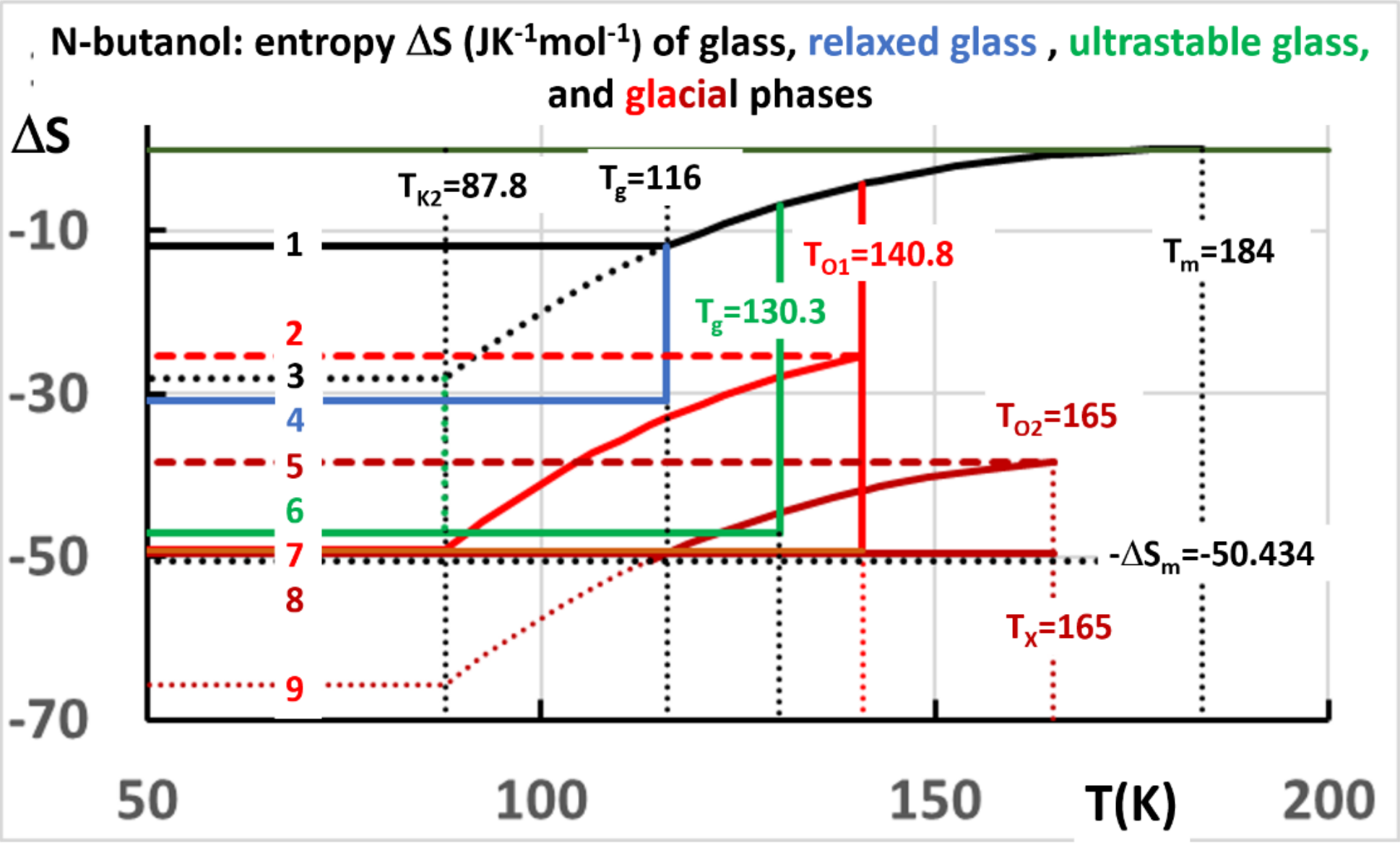

**Figure 14**: Configurational entropy of glass, ultrastable and glacial phases. 1- Black line: ΔS = (-11.904 $JK^{-1}mol^{-1}$), the glass phase; 2- Red Line: ΔS = (-25.349 $JK^{-1}mol^{-1}$), the first glacial phase below $T_{O1}$ = 140.8 K for high heating rates; 3- Black Line: ΔS = (-28.124 $JK^{-1}mol^{-1}$), the underlying and undercooled Phase 3 below $T_{K2}$ = 87.8 K; 4- Blue Line: ΔS = (-30.954 $JK^{-1}mol^{-1}$), the fully-relaxed glass up to $T_g$ = 284 K; 5- Brown Line: ΔS = (-38.372 $JK^{-1}mol^{-1}$), the second glacial phase for high heating rates; 6- Green Line: ΔS = (-47.174 $JK^{-1}mol^{-1}$), the ultrastable glass phase up to $T_g$ =130.3 K; 7- Red Line: ΔS = (-49.212 $JK^{-1}mol^{-1}$), the first glacial glass up to $T_{O1}$ = 140.8 K at low heating rate; 8- Brown Line 8: ΔS = (-49.534 $JK^{-1}mol^{-1}$) for the second glacial phase during rapid cooling and heating rates up to $T_{O2}$ = 165 K; 9- Brown line: ΔS = (-65.754 $JK^{-1}mol^{-1}$), the second glacial phase crystallizes below 116 K at low cooling and heating rates. ΔS = (-50.434 $JK^{-1}mol^{-1}$), the crystallization entropy.

The entropies of various phases represented in Figure 14 are calculated using (18) with $\varepsilon_{ls0}$, $\varepsilon_{gs0}$, $\theta_{og}^2$, and $\theta_{om}^2$ given in (28–30), substracting the entropy changes at the first-order transition temperatures $T_{O1}$ and $T_{O2}$ for the two glacial phases and at $T_{K2}$ for the ultrastable glass phase, using the calculated characteristic temperatures represented in Figure 13. The crystallization temperature occurs at $T_X = T_{n-} = T_g = 165$ K because the glass relaxes toward the crystal entropy instead of being melted. The glass transition occurs when the glass frozen entropy along Line 8 becomes equal to the entropy $\Delta S$ = (-49.53 $JK^{-1}g^{-1}$) of the second glacial phase. The crystallization temperature is in perfect agreement with the experimental observation in Figure 11. The calculated entropy is correct because it predicts the experimental value of $T_X$.

The entropy $\Delta S$ = (-30.954 $JK^{-1}mol^{-1}$) of fully-relaxed glass phase is recovered at $T_g$ = 116 K as shown in Figure 14 and is composed of the glass entropy (-11.904 $JK^{-1}mol^{-1}$) and of the relaxation entropy ($-\Delta\varepsilon_{lg0}\times\Delta H_m/T_{K2}$ = -19.050 $JK^{-1}mol^{-1}$).

The entropy of ultrastable glass phase formed at $T_{K2}$ = 87.8 K from $\Delta\varepsilon_{lg} = 0$ is recovered at $T_g$ = 130.3 K as shown in Figure 14. The first-order transition of undercooled Phase 3 at $T_{K2}$ is accompanied by an enthalpy coefficient change ($-\Delta\varepsilon_{lg0}$ = -0.1803) and by an entropy change $\Delta S$ of ($-\Delta\varepsilon_{lg0}\times\Delta H_m/T_{K2}$ = -19.050 $JK^{-1}mol^{-1}$) leading to a configurational entropy (-47.174 $JK^{-1}g^{-1}$).

The enthalpy coefficient and entropy of the glacial phase equal to (-0.66911) and (-49.534 $JK^{-1}mol^{-1}$) respectively are obtained by a slow cooling from an annealing temperature of ≅ 140 K down to $T_{K2}$ = 116 K and using a heating of 3 $Kmin^{-1}$ up to $T_X$ = 165 K. The liquid phase becomes vitreous with constant enthalpy and entropy up to $T_X$. This description still shows that denser glass phases of bulk samples can be obtained using the same process of annealing above $T_g$ in many other glasses. This analysis is now applied to $Zr_{41.2}Ti_{13.8}Cu_{12.5}Ni_{10}Be_{22.5}$ (Vit1).

## 5 Liquid–liquid first-order transitions

### 5.1 $Zr_{41.2}Ti_{13.8}Cu_{12.5}Ni_{10}Be_{22.5}$ (Vit1)

#### 5.1.1 Heat capacity, viscosity measurements, and liquid-liquid transitions of Vit1

Vit1 has a glass transition temperature $T_g$ = 625 K [41]. Its heat capacity jump at $T_g$ is $\Delta C_p (T_g)$ ≅ 21.6 $JK^{-1}$g-atom$^{-1}$ [72] and its melting heat $\Delta H_m$ = 8680 Jg-atom$^{-1}$. The shear rate and temperature dependence of the viscosity in Vit1 have been measured in the liquid and undercooled liquid states between 907 and 1300 K [40]. After quenching the alloy into a glassy state, the reheated material displays very high viscosity values above the liquidus temperature of 1026 K. With increasing temperature, a transition above 1225 K is observed with a drastic drop in viscosity that is associated with the disappearance of shear thinning. The increased viscosity is only recovered when the liquid is deeply undercooled. The shear thinning and the transition at 1225 K are attributed to the destruction of medium-and-short-range order in the liquid [40]. These phenomena have recently been restudied using calorimetric measurements and synchrotron X-ray scattering [41]. A heat capacity peak of superheated liquid after supercooling is observed during heating around T = 1116 K accompanied by an endothermic latent heat of about 1100 $Jmol^{-1}$. Structural changes corresponding to these anomalies are observed with in-situ synchrotron X-ray-scattering experiments in a contactless environment using an electrostatic levitator (ESL) as reproduced in Figure 15. This transformation is viewed as a crossover of dynamics from the strong to fragile liquid and is consistent with the observations in viscosity. There is an endothermic liquid–liquid transition during heating reinforced by the symmetrical observation of an exothermic latent heat regarding $T_m$ = 965 K and an exothermic structural change around 816 K by supercooling.

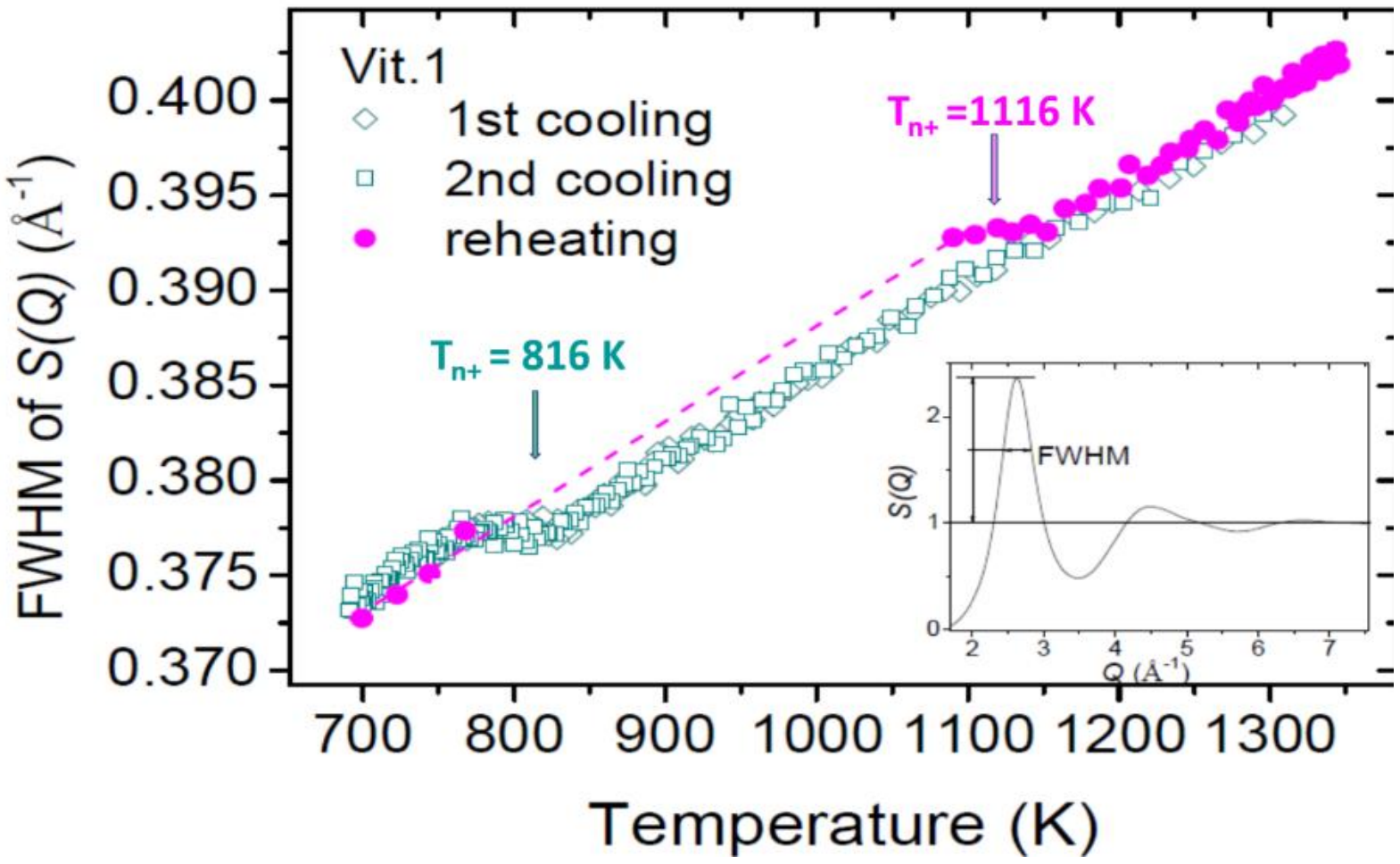

**Figure 15**: "The FWHM of the 1st peak of S(Q) (see inset) versus temperature during thermal cycles. The arrows point out the clear slope changes in the temperature range 760–830 K during cooling and 1,100–1,200 K upon reheating. The dashed line is the assumed heating data trace if crystallization can be avoided on reheating". The calculated temperatures $T_{n+}$ predicted by the model using (3) and $T_m$ = 965 K are added. This agreement with predictions confirms that $T_m$ = 965 K as suggested. Reprinted from [S. Wei, F. Yang, J. Bednarcik, I. Kaban, O. Shuleshova, A. Meyer & R. Busch, *Nature Commun.* 4 (2013), 2083].

### 5.1.2 Model predictions for Vit1

The liquid is fragile because the specific heat jump at $T_g$ is much larger than $1.5\times\Delta H_m/T_m$ = 13.5 $JK^{-1}g\text{-}atom^{-1}$ and the VFT temperature larger than $T_m/3$. Applying (12) with a = 0.833 leads to the experimental value $\Delta C_p(T_g) \cong 21.6\ JK^{-1}g.atom^{-1}$. Equations (11–15) are applied with $\theta_g$ = (-0.35308), $T_m$ = 965 K and a = 0.833 to obtain ($\varepsilon_{ls}$) of Liquid 1, ($\varepsilon_{gs}$) of Liquid 2, and ($\Delta\varepsilon_{lg}$) of Phase 3 given by (31–33)):

$$\varepsilon_{ls} = 1.70651 \times (1 - \theta^2/0.2226) \qquad (31)$$

$$\varepsilon_{gs} = 1.4715 \times (1 - \theta^2/0.34564) \quad (32)$$

$$\Delta\varepsilon_{lg} = 0.23501 - \theta^2 \times 3.409 \quad (33)$$

where $\Delta\varepsilon_{lg0} = (\varepsilon_{ls0}-\varepsilon_{gs0}) = 0.23501$ in (33). The temperatures $T_{VFT} = T_{0m} = 509.7$ K and $T_{K1} = 532$ K are deduced from (18,22,23). The enthalpy coefficient of Vit1 Phase 3 at $\theta_{0m}$ is $\Delta\varepsilon_{lg}(\theta_{0m}) = (-0.52382)$ and $\Delta\varepsilon$ in (7) is expected to be $(0.52382-\Delta\varepsilon_{lg0}) = 0.28882$ for the first-order transition to the glacial phase.

The predicted characteristic temperatures in Figure 16 are, for part of them, in agreement with the experimental observations [40,41,72]: $T_{K1} = 532$ K calculated with (18); $T_{0m} = 509.7$ K with (13); $T_{K2} = 606.7$ K with ($\Delta\varepsilon_{lg} = -\Delta\varepsilon_{lg0} = -0.23501$) in (16); $T_g = 625$ K; $T_{O1} = 834$ K, the first-order transition temperature associated with glacial phase formation using (18) and entropy considerations; $T_{Br-} = 712$ K and $T_{Br+} = 1218$ K for $\Delta\varepsilon_{lg} = 0$ in (16); $T_{gg} = 876$ K the virtual glass transition temperature predicted with $\Delta\varepsilon = 0.28882$ in (7); $T_{n+} = 816$ and 1116 K for $\Delta\varepsilon_{lg} = \theta_{n+}$ in (3). The following temperatures are not indicated in Figure 16; $T_X = 695$ K calculated with (18) using entropy considerations, $T_{gg} = 876$ K, and $T_m = 965$ K. The liquid-liquid transitions predicted at $T_{n+}$, $T_{Br+}$ and $T_X$ are in perfect agreement with the experimental values [41,73].

Vit1 liquid Phase 3 has two nucleation temperatures $T_{n+}$. The liquid order disappears at $T_{n+} = 1116$ K and reappears by Phase 3 nucleation at $T_{n+} = 816$ K. Phase 3 being the supercooling phase, the order is only observable, as shown in 2007 [40] in the viscosity by supercooling the liquid below $T_{n+} = 816$ K. The model developed here predicts $\Delta\varepsilon_{lg} = \theta_{n+} = 0.15407$ and the latent heat of the transition 1337 $Jmol^{-1}$ can be compared with the experimental value 1100 $Jmol^{-1}$. The very-high–density amorphous phase of water under pressure [35] also occurs at a temperature $T_{n+}$ above the ice melting temperature $T_m$ under pressure [21].

The enthalpy coefficient of the fully-relaxed glass phase (-0.42318) is composed of the glass enthalpy coefficient (-0.18817) and the maximum relaxation enthalpy coefficient equal to the latent heat coefficient (-0.23501) at $T_{K2}$.

The homogeneous nucleation temperatures $T_{n-}$ of superclusters in Liquid 2 are given in (7). A coefficient $\Delta\varepsilon$ equal to $\Delta\varepsilon_{lg0} = 0.23501$ leads to a first-order transition at $T_{K2}$ below $T_g$ given by (17). A coefficient $\Delta\varepsilon = 0.28882$ leads to $T_{n-} = T_{gg} = 876$ K with (7). The entropy of this new fragile liquid with $T_{gg} = 876$ K is calculated using (18) with the new coefficients $\varepsilon_{gs0} = 1.86219$, $\theta_{0g}^2 = 0.11405$, and compared with that of the initial liquid with (18). The transition at $T_{O1} = 834$ K transforms this new glacial Phase 3 in the initial Phase 3 because the associated entropy change at $T_{O1}$ between these two phases becomes equal to $(0.28882 \times \Delta H_m/T_{O1} = 3.0058$ $JK^{-1}$g-atom$^{-1}$). The enthalpy is expected to remain constant below $T_{O1}$ during a very high heating rate. ($\Delta\varepsilon_{lg}$) is represented in Figure 16 as being constant and equal to (-0.52383). It is also expected to vary during the decrease of heating rates, instead of being the same below $T_{O1}$. At very low heating rate, the enthalpy coefficient of the glacial phase would have to become constant below $T_{K2}$ = 606.7 K. This glacial phase is not observed, up to now, and is only a prediction.

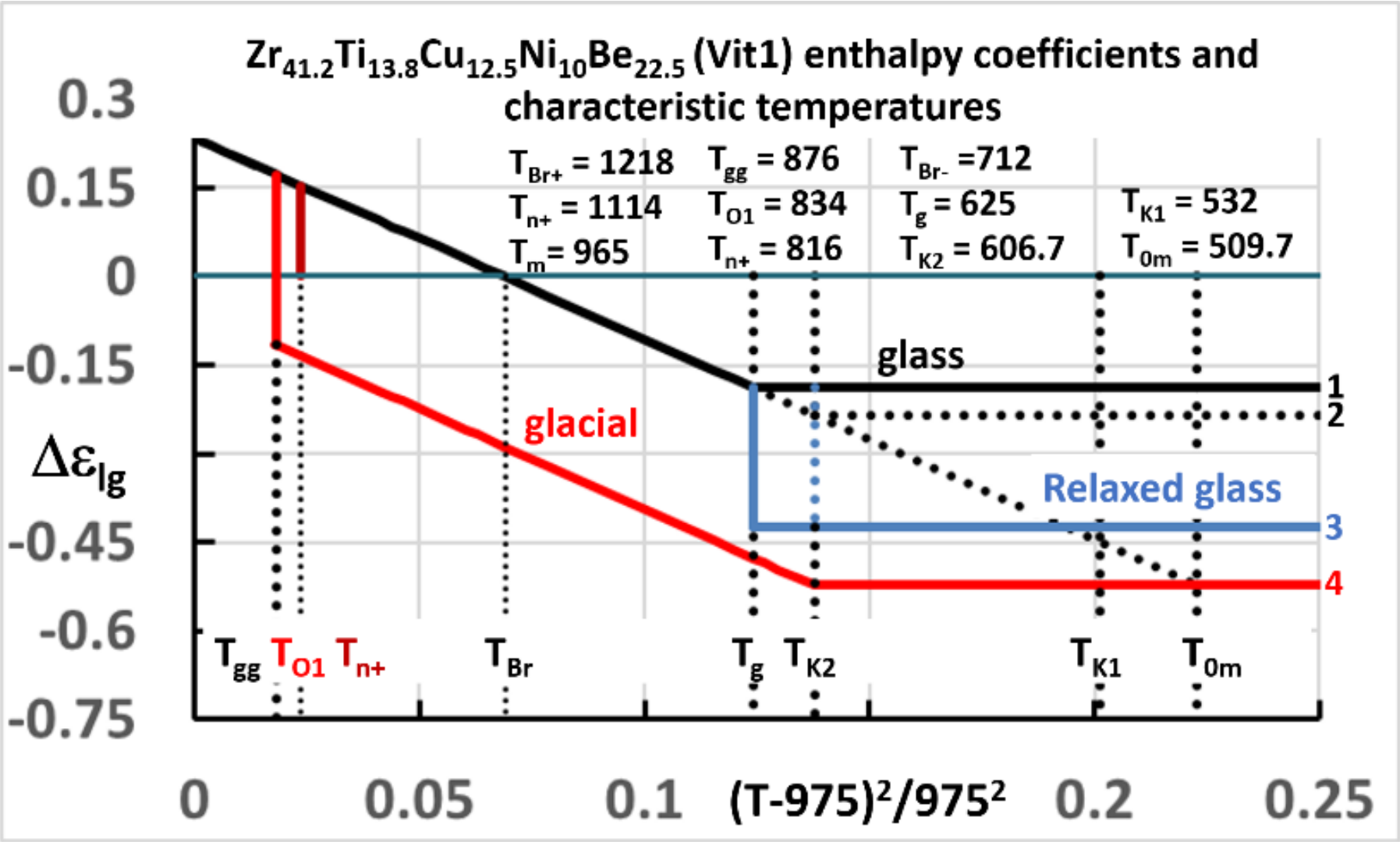

**Figure 16**: Enthalpy coefficients of Vit1 Phase 3 versus the square of the reduced temperature $\theta^2 = (T\text{-}T_m)^2/T_m^2$. Below $T_{K2}$, for underlying Phase 3, $\Delta\varepsilon_{lg}(\theta)$ = (-0.23501); for ultrastable glass $\Delta\varepsilon_{lg}(\theta)$ = (-0.47) (not represented); for glacial Phase 3, $\Delta\varepsilon_{lg}(\theta)$ = (-0.52382); for freezing of the glass, $\Delta\varepsilon_{lg}(\theta)$ = (-0.18817) below $\theta_g$; for fully-relaxed glass, $\Delta\varepsilon_{lg}(\theta)$ = -0.42318. The characteristic temperatures: $T_{gg}$ =876 K, $T_{O1}$ = 834 K, $T_{n+}$ = 816 and 1114 K, $T_{Br-}$ = 712 K, $T_{Br+}$ =1218 K, $T_g$ =625 K, $T_{K2}$ = 606.7 K, $T_{K1}$ = 532 K, $T_{0m}$ = 509.7 K.

The crystallization temperature Tx is calculated to be equal to 695 K where the entropy calculated using (18) is still available to accommodate the entropy 3.0058 $JK^{-1}g\text{-atom}^{-1}$ of the first-order transition at $T_{O1}$. This temperature is exactly equal to the experimental temperature $T_{X1}$ [73]. The ultrastable glass phase has the lowest enthalpy coefficient (-0.47 = $-2\times\Delta\varepsilon_{lg0}$) below $T_{K2}$ after a first-order transition which can be obtained after liquid hyper-quenching or by slow vapor deposition at $T_{K2}$. This phase is not represented because its entropy would lead to crystallization. This is not the case for the glacial phase.

The enthalpy coefficient of Vit1 glacial phase below $T_{K2}$ is equal to (-0.52382) which is the minimum value of $\Delta\varepsilon_{lg}$ at $T_{0m}$ as shown in Figure 16. The latent heat coefficient of Vit1 glacial phase at $T_{O1}$ is constant below the underlying first-order transition at $T_{K2.}$ The sum of the two enthalpy coefficients (-0.28882) at $T_{O1}$ and (-0.23501) at $T_{K2}$ are equal to (-0.52382). This assumption has been successfully applied to triphenyl phosphite, d-mannitol, and n-butanol. There has been no observation, up to now, of a glacial phase in Vit 1. Glacial phases would have to exist in all glass-forming melts.

The entropies ΔS in $JK^{-1}g\text{-atom}^{-1}$ of underlying glass Phase 3, fully-relaxed glass phase and glacial Phase 3 are represented in Figure 17 as a function of temperature. The frozen entropy (- 5.0315 $JK^{-1}$g-atom$^{-1}$) of the glass phase is not represented. The entropy ΔS = (-8.3939 $JK^{-1}$g-atom$^{-1}$) of fully-relaxed glass phase is recovered at $T_g$ = 625 K and is composed of the glass entropy (-5.0315 $JK^{-1}$g-atom$^{-1}$) and of the relaxation entropy ($-\Delta\varepsilon_{lg0}\times\Delta H_m/T_{K2}$ = -3.3624 $JK^{-1}$g-atom$^{-1}$). The ultrastable phase has an entropy lower than that of crystals and cannot be obtained without crystallization. The glacial phase with ΔS (θ) = (-8.6975) is more stable than the ultrastable glass phase.

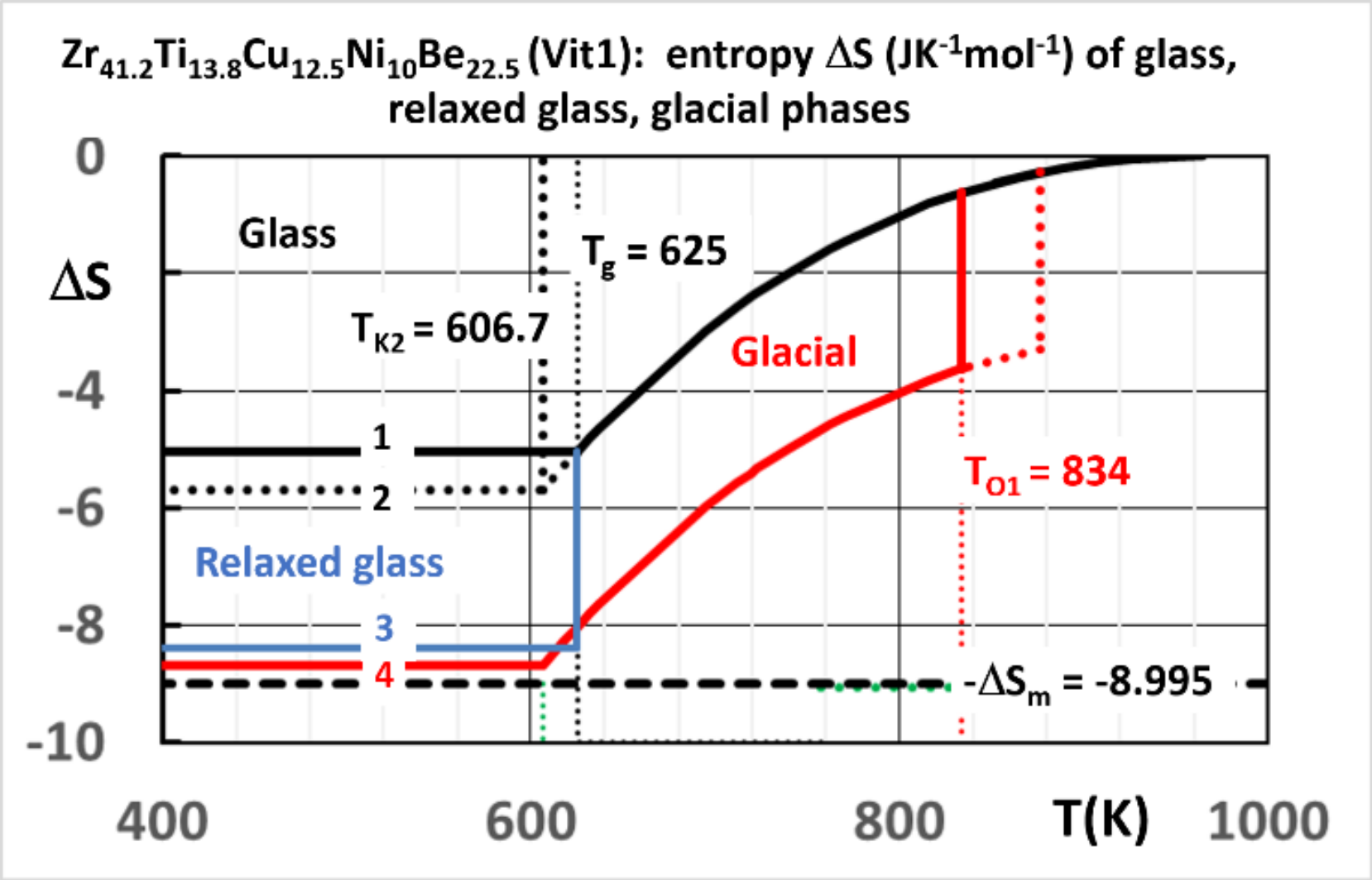

**Figure 17**: Entropy coefficients of Vit1 Phase 3 versus T (K). 1- ΔS(θ) = (-5.0315 $JK^{-1}$g-atom$^{-1}$), frozen entropy of the glass, below $\theta_g$. 2- ΔS(θ) = (-5.6917 $JK^{-1}$g-atom$^{-1}$), below $T_{K2}$ = 606.7 K, for underlying Phase 3. 3- ΔS = (-8.3939 $JK^{-1}$g-atom$^{-1}$), the fully-relaxed entropy of the glass phase. 4- ΔS(θ) = (-8.6975 $JK^{-1}$g-atom$^{-1}$) for the glacial Phase 3. Entropy of crystals equal to (-8.995 $JK^{-1}$g-atom$^{-1}$). The ultrastable glass phase crystallizes. The characteristic temperatures: $T_{gg}$ = 876 K (glacial), $T_{O1}$ = 834 K, $T_{n+}$ = 816 and 1114 K, $T_{Br-}$ = 712 K, $T_{Br+}$ = 1218 K, $T_g$ = 625 K, $T_{K2}$ = 606.7 K, $T_{K1}$ = 532 K, $T_{0m}$ = 509.7 K.

### 5.2 CoB eutectic liquid and Sn droplets

#### 5.2.1 Critical supercooling and superheating in eutectic CoB alloys and tin droplets

Cyclic superheating and cooling are carried out for the undercooled hypereutectic $Co_{80}B_{20}$, eutectic $Co_{81.5}B_{18.5}$, and hypoeutectic $Co_{83}B_{17}$ alloys [42]. For each alloy, there is a critical superheating temperature $T_c$ for which there is a sharp increase of the mean undercooling. DSC measurements above $T_m$ show that there is a corresponding small endothermic peak during heating at a temperature, nearly equal to $T_c$. An example of this work is reproduced in Figure 18. The undercooling, calculated from $T_m$ = 1406 K, is equal to 220 K for eutectic $Co_{81.5}B_{18.5}$. The endothermic heat is observed at 1656 K with $T_c$ =1653 K.

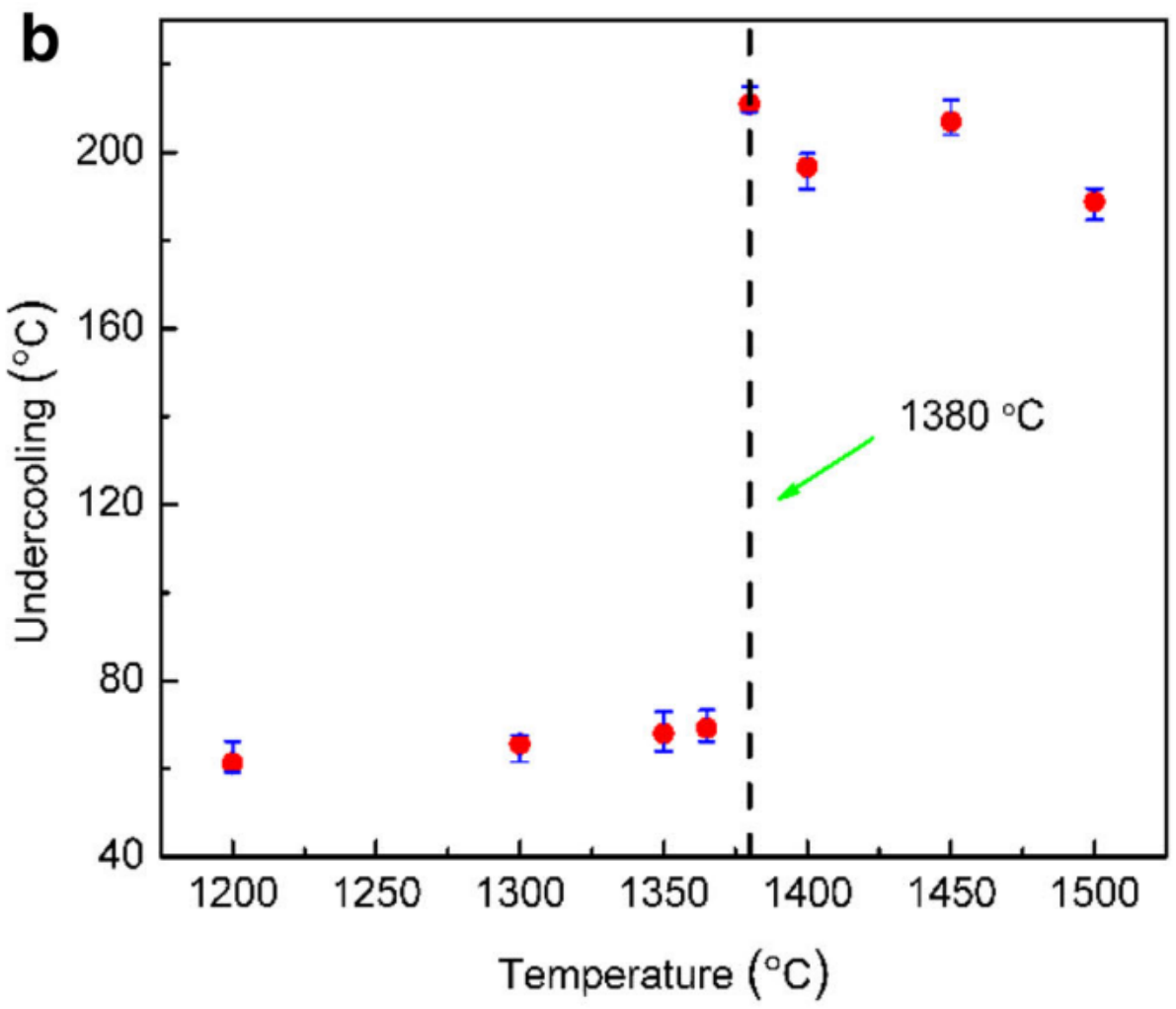

**Figure 18**: Mean undercooling of eutectic $Co_{81.5}B_{18.5}$ with different overheating temperatures. Critical overheating temperature $T_c$ equal to 1653 K above $T_m$ = 1406 K. Reprinted from [Y. He, J. Li. J. Wang, H. Kou, E. Beaugnon, *Applied Physics A.* 123 (2017) 391] with the permission of Springer Nature.

The solidification of a pure Sn single micro-sized droplet is studied by differential fast scanning calorimetry with cooling rates in the range 500 to 10 000 $Ks^{-1}$ [43]. The sample has a spherical shape covered by an oxide layer. A critical undercooling 99 ± 2 K is observed as reproduced in Figure 19 and corresponds to $\theta = (T-T_m)/T_m = 0.194$ with $T_m$ = 520 K.
In these two examples, the existence of critical superheating and supercooling related to liquid and solid nucleation critical temperatures is shown.

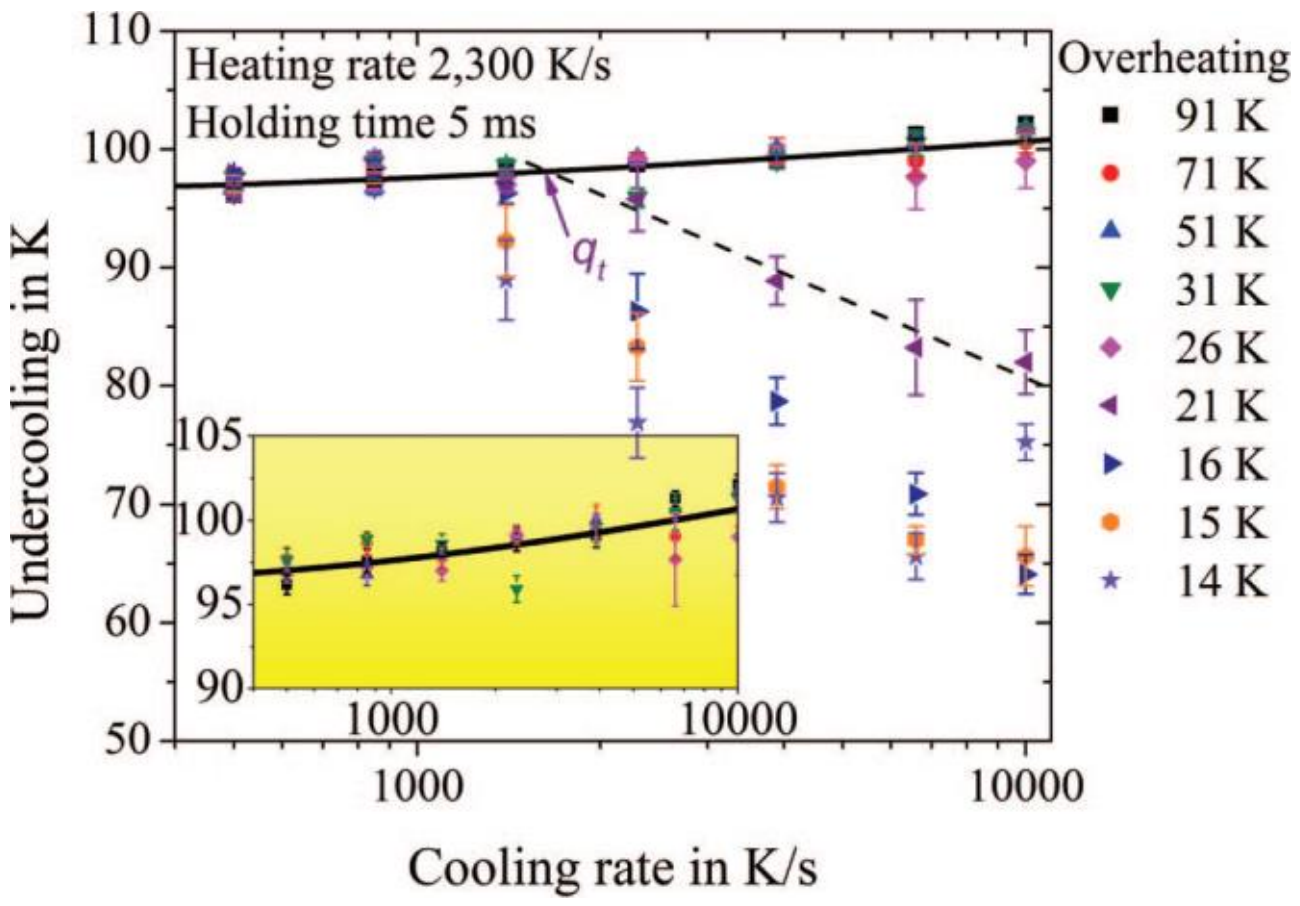

**Figure 19**: Undercooling dependence on cooling rate and overheating: solid curve – surface nucleation and dashed curve- linear fit. The inset shows a close-up of the undercooling plateau

(99 ± 2 K). Each point is the average of 10 identical measurements and the error bars are the standard error of 10 identical measurements for each temperature course. Reprinted from [B. Yang, J.H. Perepezko, J.W.P. Schmelzer, Y. GaO, and C. Schick, *J. Chem. Phys.* 140 (2014) 104513] with the permission of AIP Publishing.

### 5.2.2 Model predictions for the critical temperatures of liquid and crystal nucleation

Homogeneous nucleation for melting in superheated crystals is used to derive a stability limit for the crystal lattice above its equilibrium melting point. It is known that at a critical temperature which is about $1.2\times T_m$ for various elemental metals, a massive homogeneous nucleation of melting occurs in the superheated crystal protected against surface melting [44]. The temperature $\theta_{n+} = \Delta\varepsilon_{lg}$ is equal to zero for all liquid elements. The second melting phenomenon due to superclusters (called "tiny crystals" in [10]) is predicted in all pure liquid elements at $\theta = 0.198$ with (3) and $\varepsilon_{ls} = \theta$ leading to $0.217\times(1-\theta^2\times2.25) = \theta$, ($\theta_{0m} = -2/3$) [10]. An undercooling limit of 99 ±2 K is observed for micro-sized droplets of Sn using overheating up to 91 K and cooling rates up to $10^4$ $Ks^{-1}$. The undercooling limit is predicted with (3) as occurring at $\theta = (-0.198)$ (102.9 K) with $T_m = 520$ K and corresponds to the value observed with the highest cooling rate.
For CoB eutectic alloys, the authors [42] use this theoretical limit to explain the melting of growth nuclei at $T_c$. The temperature $\theta_{n+}$ is in fact equal above $T_m$ to $\theta_c = +0.1756$. This limit $\theta_{n+} = \theta_c$ for $Co_{81.5}B_{18.5}$ is obtained by considering the melting temperature of ordered Phase 3 and calculating the glass transition temperature of a strong liquid with (34–36) for $\varepsilon_{ls}$, $\varepsilon_{gs}$ and $\Delta\varepsilon_{lg}$ with $\theta_{0m}^{-2} = 1.667$ ($T_{0m} = 317$ K, approximating a quasi-Arrhenius law for the high temperature viscosity):

$$\varepsilon_{ls} = 0.79926 \times (1 - \theta^2 \times 1.667) \quad (34)$$

$$\varepsilon_{gs} = 0.60103 \times (1 - \theta^2) \quad (35)$$

$$\Delta\varepsilon_{lg} = 0.19823 - \theta^2\times0.73134 \quad (36)$$

The characteristic temperatures represented in Figure 20 are: $T_g = 674$ K, $\theta_g = -0.52063$, $T_{Br+} = 2138$ K, $\theta_{Br+} = 0.52063$, $T_{n+} = 1653$ K $= T_c$, $\theta_{n+} = 0.1756$, $T_m = 1406$ K, $\theta_m = 0$, $T_{n+} = 1159$ K, $\theta_{n+} = (-0.1756)$, $T_{K2} = 371$ K, $\theta_{K2} = (-0.73628)$, $T_{K1} = 330$ K, $\theta_{K1} = (-0.76515)$, $T_{0m} = 317$ K, $\theta_{0m} = (-0.7745)$. The calculated glass transition temperature $T_g = 674$ K can be compared with those of $Fe_{84}B_{16}$ (675 K) [74] and $Fe_{73}Co_{12}B_{15}$ (685 K) [75] which are equal to the crystallization temperature $T_{X1}$ measured with a heating rate of 20 and 40 $Ks^{-1}$.

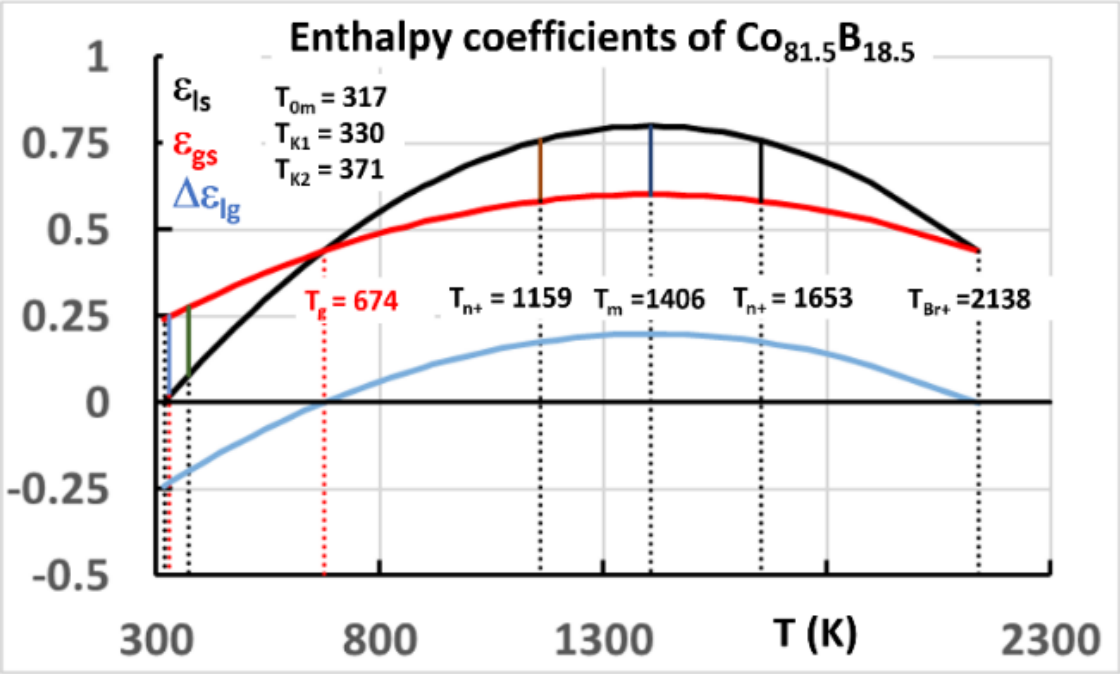

**Figure 20**: Enthalpy coefficients of amorphous $Co_{81.5}B_{18.5}$ alloy undergoing a liquid-liquid transition at $\theta_{n+} = \pm\ 0.1756$. Characteristic temperatures: $T_{0m}$ =317 K, $T_{K1}$ = 330 K, $T_{K2}$ = 371 K, $T_g$ = 674 K, $T_{n+}$ = 1159 and 1653 K ($\theta_{n+} = \pm\ 0.1756$), $T_m$ = 1406 K ($\theta = 0$), $T_{Br+}$ =2138 K ($\theta_{Br+}$ = 0.5206).

### 5.3 Bulk metallic glass $Ti_{34}Zr_{11}Cu_{47}Ni_8$

#### 5.3.1 Heat capacity, undercooling versus overheating and recalescence

This bulk metallic glass has the following properties: the glass transition temperature $T_g$ = 671 K, the melting temperature $T_m$ = 1150 K, the melting heat, $\Delta H_m$ =11300 $JK^{-1}$g-atom$^{-1}$, the heat capacity jump at $T_g$, $\Delta C_p$ ($T_g$) = 1.5×$\Delta H_m/T_m$ =14.7 $JK^{-1}$g-atom$^{-1}$ [76,77]. Using electrostatic levitation, undercoolings of 226.3 K are observed after specimen overheats of 300 K [39]. After undercooling, crystallization and recalescence, the temperature increases up to ≅ 1150 K except for the undercooling resulting from overheats of 300 K where the temperature is 43 K lower as reproduced in Figure 21. A fusion enthalpy of 1850 Jg-atm$^{-1}$ is missing which is calculated with the mean heat capacity of the specimen [77] of 43 $JK^{-1}$g-atom$^{-1}$ at these temperatures. This phenomenon is considered by the authors as the formation of a metastable phase before crystallization during cooling which is still supported by a consistent deviation in the cooling curve at 961 K. This temperature is 199 K below $T_{liq}$ = 1160 K.

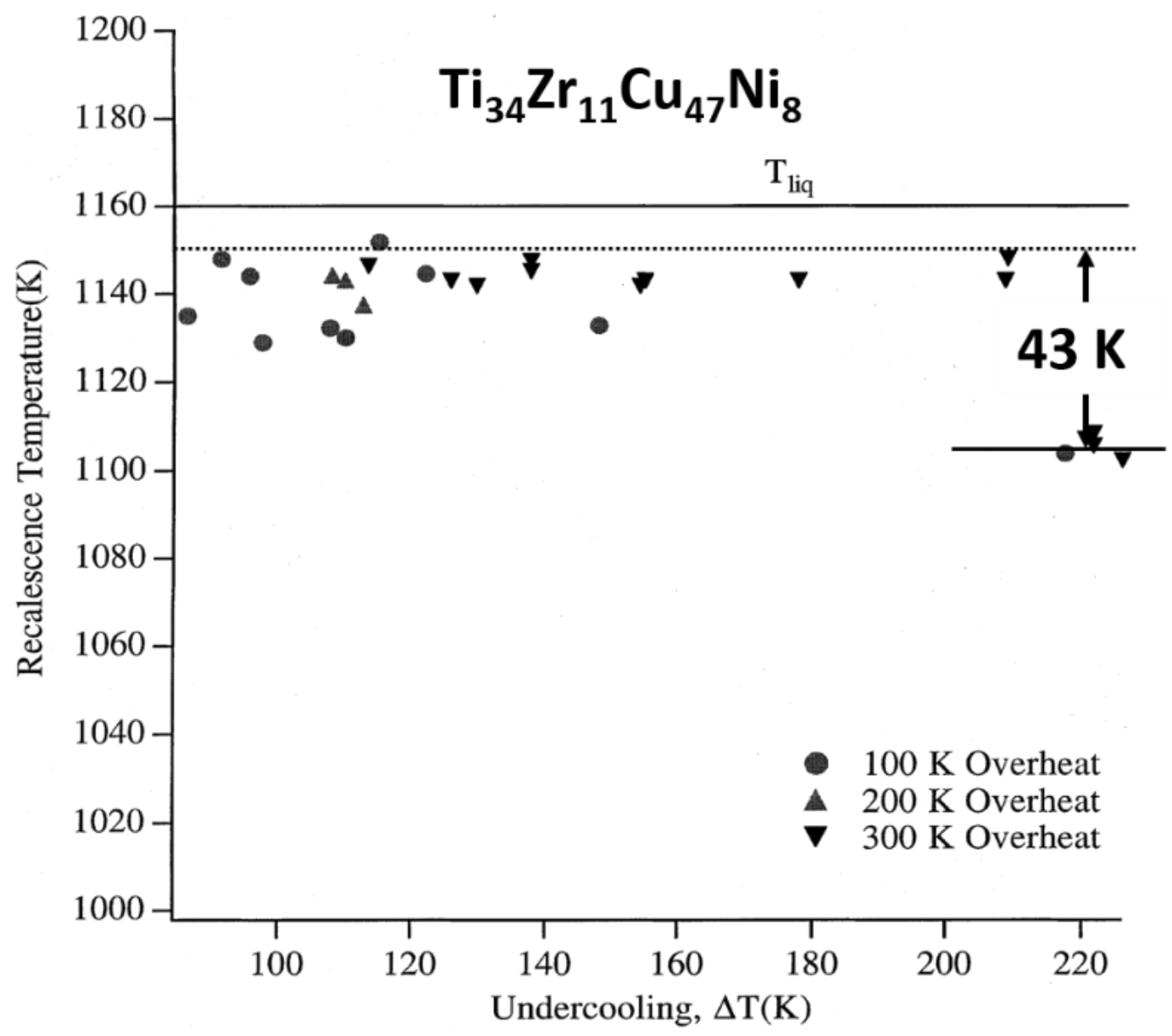

**Figure 21**: The recalescence temperature $T_{rec}$ is plotted as a function of undercooling ΔT [39]. The recalescence temperatures at high overheating rates do not reach the line of the maximum recalescence temperatures of ≅ 1150 K. The data shows a dramatic decrease in the recalescence temperature applying an overheat of 300 K leading to undercoolings of 220 K. Reprinted from [C.C. Hays, W.L. Johnson, *J. Non-Cryst. Sol.* 250-252 (1999) 596-600] with the permission of Elsevier.

### 5.3.2 Model predictions for critical temperatures and latent heats of ordered liquid Phase 3

The liquid is fragile because the specific heat jump at $T_g$ is equal to $1.5\times\Delta H_m/T_m = 14.7$ $JK^{-1}$g-atom$^{-1}$. Equations (11–15) are used with $\theta_g = -0.41652$, $T_m = 1150$ K, a = 1 to obtain ($\varepsilon_{ls}$) of Liquid 1, ($\varepsilon_{gs}$) of Liquid 2, and ($\Delta\varepsilon_{lg}$) of Phase 3 given in (37-39)):

$$\varepsilon_{ls} = 1.58348 \times (1 - \theta^2/0.29313) \quad (37)$$

$$\varepsilon_{gs} = 1.37522 \times (1 - \theta^2/0.38187) \quad (38)$$

$$\Delta\varepsilon_{lg} = 0.20826 - \theta^2\times 1.8007 \quad (39)$$

The predicted characteristic temperatures are: $T_{0m} = 527$ K with (13); $T_{K2} = 597$ K with $\Delta\varepsilon_{lg} =$ ($-\Delta\varepsilon_{lg0} = -0.20826$) in (16); $T_g = 671$ K; $T_{Br-} = 759$ K and $T_{Br+} = 1541$ K for $\Delta\varepsilon_{lg} = 0$ in (16); $T_{n+} =$ 964 and 1336 K for $\Delta\varepsilon_{lg} = \theta_{n+} = (\pm\ 0.16137)$ in (3).

The calculated temperature $T_{n+} = 964$ K is in very good agreement with the experimental observations of a residual transition at 961 K [39]. The crystallization starts below 961 K and produces recalescence removing the latent heat of Phase 3 which is ordered up to $T_{n+}$ =1336 K. The predicted latent heat of Phase 3 is 0.16137×$\Delta H_m$ = 1823 Jg-atom$^{-1}$, in agreement with the missing fusion heat of 1850 Jg-atom$^{-1}$. This missing fusion enthalpy still shows the existence of a first-order transition at the nucleation temperature 964 K of the ordered liquid Phase which is recovered at its melting temperature $T_{n+}$ =1336 K. Then, the total melting enthalpy remains equal to $\Delta H_m$.

**Conclusions**:

The classical nucleation equation completed with an additional enthalpy $\varepsilon_{ls}$ ×$\Delta H_m$ equal to $\varepsilon_{ls0}$×$\Delta H_m$ at $T_m$ depending on $\theta^2 = (T-T_m)^2/T_m^2$ predicts the formation and the melting of critical superclusters in Liquid 1 at three homogeneous nucleation temperatures with two below $T_m$ and one above $T_m$. These critical superclusters are so numerous at these temperatures that they occupy all the liquid volume because their nucleation rate is equal to one. An ordered liquid is built during cooling the initial homogeneous Liquid 1 at the lowest nucleation temperature $T_1$ in the no man's land, where crystallization occurs. The second nucleation temperature $T_{n+} < T_m$ above $T_1$ gives rise to solid superclusters and the third one is the melting of these solid entities by heating the ordered liquid up to the third homogeneous nucleation temperature $T_{n+}$ above $T_m$. These temperatures are observed in pure liquid elements by studying undercooling as a function of overheating. Critical overheating and undercooling are associated with these two temperatures $T_{n+}$. A first-order transition of crystal melting at $T_{n+}$ is known for a crystal protected against surface melting by a solid cover. Some residual superclusters are still melted at $T_{n+}$ even after melting crystals at $T_m$. The homogeneous nucleation temperature $T_1$ is an ordering temperature without freezing in the growth of initial superclusters formed at $T_{n+}$ below $T_m$.

Liquid 2 has an enthalpy saving coefficient $\varepsilon_{gs}$ (equal to $\varepsilon_{gs0}$ at $T_m$). The new glass phase is formed at a homogeneous nucleation temperature of Liquid 2 by mixing the ordered states of Liquid 1 and Liquid 2 and yielding a frozen microstructure of touching and interpenetrating superclusters as shown by numerical simulations. The VFT temperature $T_{0g}$ governing the relaxation time and the viscosity is smaller than that of Liquid 1 and leads to an enthalpy coefficient difference equal to $(\varepsilon_{ls}-\varepsilon_{gs})$ below $T_g$ and a heat capacity jump at $T_g$. A new undercooled liquid phase, that I call Phase 3, is formed with this enthalpy difference between Liquid 1 and Liquid 2 equal to $\Delta\varepsilon_{lg}$ ×$\Delta H_m$= $(\varepsilon_{ls0}-\varepsilon_{gs0})$×$\Delta H_m$ and with three new nucleation reduced temperatures: $\theta_g$ and $\theta_{n+} = \Delta\varepsilon_{lg}$ (being positive or negative). Phase 3 is ordered up to its melting temperature $T_{n+}$ above $T_m$ and reappears by cooling below $T_{n+} < T_m$. It has a melting heat coefficient equal to $\Delta\varepsilon_{lg0} = (\varepsilon_{ls0}-\varepsilon_{gs0})$. The enthalpy difference between Liquid 1 and Liquid 2 cannot be larger than $\Delta\varepsilon_{lg0}$×$\Delta H_m$. An underlying first-order transition without latent heat occurs at the temperature $T_{K2}$ where $\Delta\varepsilon_{lg}(\theta_{K2})$ of Phase 3 is equal to the limit $(-\Delta\varepsilon_{lg0})$ despite the fact that Phase 3 is supplanted by the glass phase below $T_g$.

Hyper-quenching the melt below $T_{K2}$ freezes an enthalpy excess which cannot be higher than $\varepsilon_{ls0}\times\Delta H_m$. A study of seven measurements of the enthalpy recovery after quenching published by several authors confirms the existence of this limit and then, of the underlying first-order transition. The temperature where the enthalpy excess begins to be recovered is predicted in agreement with experiments by using the new nucleation temperature modified by the presence of the enthalpy excess equal to $\Delta\varepsilon_{lg0}\times\Delta H_m$.

The enthalpy excess is obtained without undergoing any glass transition during a very short time at the hyper-quenching temperature. A first-order transition leading to the ultra-stable Phase 3 is predicted at the temperature $T_{K2}$ where the total enthalpy coefficient is initially equal to zero and the relaxation time is expected to be small. This sharp transition producing latent heat equal to $2\times\Delta\varepsilon_{lg0}\times\Delta H_m$ and inducing recalescence is difficult to realize without a very-efficient thermal exchange at $T_{K2}$. The ultrastable glass transition temperature is larger than that of the glass phase for the same heating rate.

The glacial phases are driven by the enthalpy and mainly by the entropy of Phase 3. The enthalpy coefficient $\varepsilon_{ls}$ being equal to zero at the VFT temperature $T_{0m}$ of Liquid 1, $\Delta\varepsilon_{lg}$ has a lower limit equal to the coefficient ($-\varepsilon_{gs}$) at this temperature. The enthalpy coefficient of the glacial phase is well-defined and equal to $\Delta\varepsilon_{lg}(\theta_{0m})$ leading to a true glass phase. The first-order transition of the glacial phase is associated with a latent heat equal to ($\Delta\varepsilon_{lg}(\theta_{0m})-\Delta\varepsilon_{lg0}$). The sum of two first-order latent heats of ultrastable and glacial phases is equal in all examples to $\Delta\varepsilon_{lg}(\theta_{0m})$. Larger values of the latent heat coefficient are obtained using longer annealing times. They could correspond to mixtures of crystalline and glacial phases because their entropies are too close to those of crystals.

Known glass and glacial phases are analyzed within this model. The major thermodynamic properties and transition temperatures of Triphenyl phosphite, D-mannitol, and n-butanol are predicted in very good agreement with observations. The characteristic temperatures of $Zr_{41.2}Ti_{13.8}Cu_{12.5}Ni_{10}Be_{22.5}$ (Vit1), $Ti_{34}Zr_{11}Cu_{47}Ni_8$, and $Co_{81.5}B_{18.5}$ are determined. The calculated first-order transition temperatures of nucleation and melting of their ordered liquid Phase 3 below and above $T_m$ are equal to the experimental values. The critical undercooling of Sn droplets equal to $\Delta T/T_m = 0.19$ corresponds to the theoretical value 0.198 predicted for all liquid elements by this model.

All these surprising results of modeling, predicting new glass phases and LLPT, are obtained with a classical nucleation equation completed by an additional enthalpy associated with solid supercluster formation submitted to Laplace pressure in melts. It was shown, many years ago, that all liquids contain intrinsic solid nuclei above $T_m$ that control solidification and magnetic texturing during cooling. Glass and liquid-liquid transitions are evidently not crystalline transitions and are governed by critical supercluster nuclei acting as building blocks of solid amorphous and liquids that are not subjected to surface melting. These stable entities are tied to the physics and chemistry of superclusters viewed as super-atoms. "The term superatom is attributed to nanoscale collections of atoms that behave as a single unit or a quantized building block by exhibiting unique shell filling, electronic or combining behavior that is reminiscent of individual atoms” [78,79].

**Acknowledgments**: A first version of this work has been presented at various conferences. Thanks to the organizers of: « Material Analysis and Processing in Magnetic fields », june 2018, 26-29 th, Grenoble, « Semiconductors, Optoelectronics and Nanostructures » ICSON 2018 Conference Paris August 20-21, « Journées de la Matière Condensée » JMC 2018 Conference Grenoble August 27-31. These talks are registered in Researchgate with DOI: 10.13140/RG.2.2.10045.92645.